\documentclass[%
 reprint,
 amsmath,amssymb,
 aps,
nofootinbib]{revtex4-1}

\usepackage{lipsum}
\usepackage{graphicx}

\usepackage[utf8]{inputenc}
\usepackage{tikz}
\usepackage{graphicx}
\usepackage{dcolumn}
\usepackage{bm}

\def\be{\begin{equation}}
\def\bea{\begin{eqnarray}}
\def\ee{\end{equation}}
\def\eea{\end{eqnarray}}
\def\bb{\bigskip}
\def\mm{\medskip}

\def\h{{1\over 2}}
\def\t{\tilde}
\def\r{\rightarrow}
\def\nn{\nonumber\\}

\begin{document}

\preprint{APS/123-QED}

\title{Resolving the black hole causality paradox}

\author{ Samir D. Mathur}
 \affiliation{%
Department of Physics, The Ohio State University, Columbus,
OH 43210, USA\\mathur.16@osu.edu\\}%


\begin{abstract}

The black hole information paradox is really a combination of two problems: the causality paradox and the entanglement problem. The causality paradox arises because in the semiclassical approximation infalling matter gets causally trapped inside its own horizon; it is therefore unable to send its information back to infinity if we disallow propagation outside the light cone. We show how the causality paradox is  resolved in the fuzzball paradigm. One needs to distinguish between two kinds of Rindler spaces: (a)  Rindler space obtained by choosing accelerating coordinates in Minkowski space and (b)   `pseudo-Rindler' space, which describes the region near the surface of a fuzzball. These two spaces differ in their vacuum fluctuations. While low energy waves propagate the same way on both spaces, infalling objects with energies $E\gg T$ suffer an `entropy enhanced tunneling' in the pseudo-Rindler spacetime (b); this leads to the nucleation of a fuzzball before the infalling object gets trapped inside a horizon. 

\end{abstract}

\maketitle


\section{\label{sec:1}Black hole puzzles}

Consider a spherical shell of mass $M$ collapsing to form a black hole. In the semiclassical approximation we find that the shell passes through its horizon at $r_h=2GM$, and ends at a singularity at $r=0$. Hawking found that the vacuum around the horizon is unstable, and leads to the creation of particle pairs \cite{hawking}. One member of the pair (carrying a net negative energy) falls into the hole and reduces its mass, while the other escapes to infinity as `Hawking radiation'. While overall energy is conserved, there are two fundamental problems with this evaporation process:

\mm

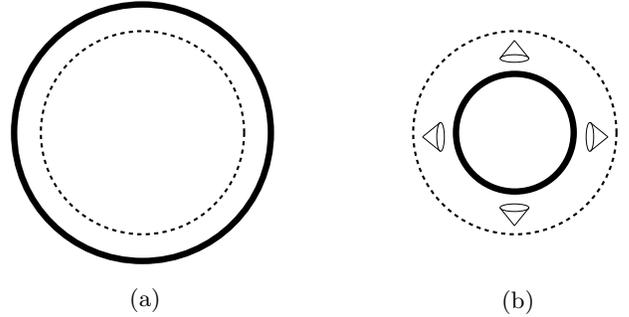
\begin{figure}

\begin{tikzpicture}[y=0.80pt, x=0.80pt, yscale=-.126000000, xscale=.126000000, inner sep=0pt, outer sep=0pt]
  \begin{scope}[shift={(0,-48.57141)}]
    \path[draw=black,dash pattern=on 1.60pt off 1.60pt,line join=miter,line
      cap=butt,miter limit=4.00,even odd rule,line width=0.800pt]
      (368.5714,580.9336) circle (10.7244cm);
    \path[draw=black,line join=miter,line cap=butt,miter limit=4.00,even odd
      rule,line width=2.400pt] (368.5714,580.9336) circle (6.2089cm);
    \path[draw=black,line join=miter,line cap=butt,miter limit=4.00,even odd
      rule,line width=0.283pt] (367.1428,302.1688) ellipse (1.5232cm and 0.3993cm);
    \path[draw=black,line join=miter,line cap=butt,miter limit=4.00,even odd
      rule,line width=0.283pt] (313.9304,300.1467) -- (367.5256,234.0937) --
      (367.5256,234.0937) -- (367.5256,234.0937) -- (367.5256,234.0937);
    \path[draw=black,line join=miter,line cap=butt,miter limit=4.00,even odd
      rule,line width=0.283pt] (421.0006,300.8207) -- (367.4054,234.7677) --
      (367.4054,234.7677) -- (367.4054,234.7677) -- (367.4054,234.7677);
    \path[xscale=1.000,yscale=-1.000,draw=black,line join=miter,line cap=butt,miter
      limit=4.00,even odd rule,line width=0.283pt] (367.1428,-862.5269) ellipse
      (1.5232cm and 0.3993cm);
    \path[draw=black,line join=miter,line cap=butt,miter limit=4.00,even odd
      rule,line width=0.283pt] (313.9304,864.5489) -- (367.5256,930.6019) --
      (367.5256,930.6019) -- (367.5256,930.6019) -- (367.5256,930.6019);
    \path[draw=black,line join=miter,line cap=butt,miter limit=4.00,even odd
      rule,line width=0.283pt] (421.0006,863.8749) -- (367.4054,929.9279) --
      (367.4054,929.9279) -- (367.4054,929.9279) -- (367.4054,929.9279);
    \path[cm={{0.0,-1.0,1.0,0.0,(0.0,0.0)}},draw=black,line join=miter,line
      cap=butt,miter limit=4.00,even odd rule,line width=0.283pt]
      (-596.6335,89.8208) ellipse (1.5232cm and 0.3993cm);
    \path[draw=black,line join=miter,line cap=butt,miter limit=4.00,even odd
      rule,line width=0.283pt] (87.7988,649.8459) -- (21.7458,596.2507) --
      (21.7458,596.2507) -- (21.7458,596.2507) -- (21.7458,596.2507);
    \path[draw=black,line join=miter,line cap=butt,miter limit=4.00,even odd
      rule,line width=0.283pt] (88.4728,542.7757) -- (22.4198,596.3709) --
      (22.4198,596.3709) -- (22.4198,596.3709) -- (22.4198,596.3709);
    \path[cm={{0.0,-1.0,-1.0,0.0,(0.0,0.0)}},draw=black,line join=miter,line
      cap=butt,miter limit=4.00,even odd rule,line width=0.283pt]
      (-596.6335,-650.1790) ellipse (1.5232cm and 0.3993cm);
    \path[draw=black,line join=miter,line cap=butt,miter limit=4.00,even odd
      rule,line width=0.283pt] (652.2010,649.8459) -- (718.2540,596.2507) --
      (718.2540,596.2507) -- (718.2540,596.2507) -- (718.2540,596.2507);
    \path[draw=black,line join=miter,line cap=butt,miter limit=4.00,even odd
      rule,line width=0.283pt] (651.5270,542.7757) -- (717.5800,596.3709) --
      (717.5800,596.3709) -- (717.5800,596.3709) -- (717.5800,596.3709);
  \end{scope}
  \begin{scope}[shift={(2.85714,-40.0)}]
    \path[draw=black,dash pattern=on 1.60pt off 1.60pt,line join=miter,line
      cap=butt,miter limit=4.00,even odd rule,line width=0.800pt]
      (-1031.4286,572.3622) circle (10.7244cm);
    \path[draw=black,line join=miter,line cap=butt,miter limit=4.00,even odd
      rule,line width=2.400pt] (-1031.4286,572.3622) circle (13.5467cm);
  \end{scope}
  \path[fill=black,line join=miter,line cap=butt,line width=0.800pt]
    (-1077.4313,1204.3622) node[above right] (text4470) {(a)};
  \path[fill=black,line join=miter,line cap=butt,line width=0.800pt]
    (319.0280,1212.9629) node[above right] (text4474) {(b)};

\end{tikzpicture}

\caption{(a) A shell of mass $M$ is collapsing towards its horizon. (b) If the shell passes through its horizon, then the information it carries is trapped inside the horizon due to the structure of light cones.} \label{fig1}

\end{figure}

(A) {\it The causality paradox:}  After the shell passes through its horizon, light cones in the region between the shell and the horizon `point inwards' as shown schematically in fig.\ref{fig1}. If we assume that we do not have any `faster than light' propagation in our theory, then the  information in the shell is causally trapped inside the horizon. Thus this information cannot escape to infinity as the hole evaporates away. What happens to this information at the endpoint of evaporation?

\mm

(B) {\it The entanglement problem:} The process of Hawking radiation creates entangled pairs at the horizon; thus we find a monotonically increasing entanglement between the radiation near infinity and the remaining hole.  Hawking's original computation was done at leading order in the semiclassical approximation, but the small corrections theorem \cite{cern} shows that this monotonic increase cannot be overcome by any source of small corrections  to the pair creation process.  What happens to this large entanglement near the endpoint of evaporation?

\mm

These two problems together make up the black hole information paradox.

One proposal to resolve both of these problems is the idea of `remnants': the evaporation process stops due to quantum gravity effects when the hole reaches planck size. The information in the infalling matter and in the negative energy members of the created pairs is then locked inside this planck mass remnant.  But string theory does not allow remnants if we accept AdS/CFT duality \cite{adscft}. Remnants must have an infinite degeneracy within an energy range $E<E_0\sim  m_p$.  On the other hand the dual CFT lives on a finite volume space $S^d$, and the CFT in a finite volume can only have a finite number of states for $E<E_0$. 
 
 Further,  there is no clear evidence that string theory allows propagation faster than the speed of light.  It is true that the theory has extended objects like strings,  but  this does not imply acausality: if we excite one end of a string, the information of this excitation travels along the string at a speed less than or equal to the speed of light. 
 
 What then is the resolution of the above puzzles in string theory? Extensive 
work in constructing black hole microstates had led to the {\it fuzzball paradigm}. For our present discussion, the relevant features of this paradigm are as follows:

\mm

(1) The microstates describing the black hole do not have a traditional horizon, i.e. there is no formation of a closed trapped surface. There is no singularity either; instead we have horizon sized quantum objects called {\it fuzzballs} whose states we write as $|F_i\rangle$ \cite{lm4,fuzzballs}. These fuzzballs radiate from their surface like normal warm bodies so there is no entanglement puzzle (B). The {\it rate} of radiation turns out to agree with that expected for Hawking radiation from individual microstates, but this radiation does not arise from pair creation since there is no region `interior to the horizon' where negative energy particles can exist \cite{radiation}.  

\mm

(2) The semiclassical collapse of a shell suggests that it passes through $r=r_h$ and a horizon does form. But as the shell reaches $r\approx r_h$ the semiclassical approximation is violated by an {\it entropy enhanced tunneling} into the space of fuzzballs $|F_i\rangle$. The probability for the collapsing shell to tunnel into any of the fuzzball states is small, as expected for transitions between two macroscopic objects:
\be
{\cal P} \sim e^{-2S_{cl}}
\ee
where $S_{cl}$ is the classical action for the tunneling process.
 But this smallness is offset by the large degeneracy
 \be
 {\cal N} \sim e^{S_{bek}}
 \ee
 of fuzzball states, where $S_{bek}$ is the Bekenstein entropy \cite{tunnel,kraus,puhm}.  As a result the shell state $|S\rangle$  transitions to a linear combination of the $|F_i\rangle$, and we then get unitarity preserving radiation just as we would get from any other warm body. 

\mm

In this paper our goal is to obtain  a picture of how, when and were the entropy enhanced tunneling should happen.  Our principal tool will be the causality paradox (A):  we require that the tunneling  happens in a way that information of a collapsing object never gets trapped inside its own horizon. Since horizons form over timescales of order the crossing time $\sim M$, this requirement provides a much stronger constraint than the  entanglement problem (B): the entangled pairs are produced over the much longer Hawking evaporation timescale $\sim M^3$, and a transition to fuzzballs over any timescale $\lesssim M^3$ would suffice to remove the entanglement problem.

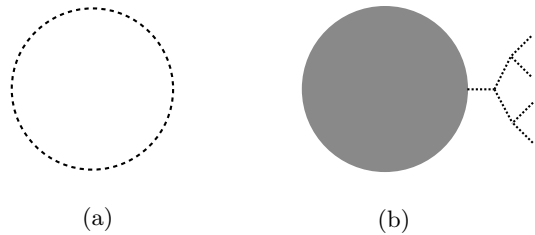
\begin{figure}

\definecolor{c898989}{RGB}{137,137,137}

\begin{tikzpicture}[y=0.80pt, x=0.80pt, yscale=-.1000000, xscale=.1000000, inner sep=0pt, outer sep=0pt]
  \path[draw=black,dash pattern=on 1.60pt off 1.60pt,line join=miter,line
    cap=butt,miter limit=4.00,even odd rule,line width=0.800pt]
    (-1028.5714,530.9337) circle (10.7244cm);
  \path[fill=black,line join=miter,line cap=butt,line width=0.800pt]
    (-1077.4313,1204.3622) node[above right] (text4470) {(a)};
  \path[fill=black,line join=miter,line cap=butt,line width=0.800pt]
    (319.0280,1212.9629) node[above right] (text4474) {(b)};
  \path[fill=c898989,dash pattern=on 1.60pt off 1.60pt,line join=miter,line
    cap=butt,miter limit=4.00,even odd rule,line width=0.800pt]
    (354.2859,530.9337) circle (11.0470cm );
  \path[draw=black,dash pattern=on 0.80pt off 0.80pt,line join=miter,line
    cap=butt,miter limit=4.00,even odd rule,line width=0.800pt]
    (871.9681,532.3266) -- (957.7462,692.3978);
  \path[draw=black,dash pattern=on 0.80pt off 0.80pt,line join=miter,line
    cap=butt,miter limit=4.00,even odd rule,line width=0.800pt]
    (1045.9737,281.0819) -- (949.5567,376.6479) -- (1045.9737,472.2140);
  \path[draw=black,dash pattern=on 0.80pt off 0.80pt,line join=miter,line
    cap=butt,miter limit=4.00,even odd rule,line width=0.800pt]
    (744.0000,532.3622) -- (869.7143,532.3622);
  \path[draw=black,dash pattern=on 0.80pt off 0.80pt,line join=miter,line
    cap=butt,miter limit=4.00,even odd rule,line width=0.800pt]
    (872.0000,532.3622) -- (952.0000,372.3622);
  \path[draw=black,dash pattern=on 0.80pt off 0.80pt,line join=miter,line
    cap=butt,miter limit=4.00,even odd rule,line width=0.800pt]
    (1054.5451,595.3676) -- (958.1281,690.9336) -- (1054.5451,786.4997);

\end{tikzpicture}

\caption{(a) The traditional black hole. The region around the horizon is locally just like empty Minkowski space, so it has the same vacuum fluctuations as empty space. (b) A fuzzball has a surface at $r=r_b=2GM+\epsilon$. The presence of this fuzzball boundary at $r=r_b$ can lead  to new virtual effects indicated by the dashed lines.} \label{fig1p}

\end{figure}

 \begin{figure}
 \includegraphics[scale=.12] {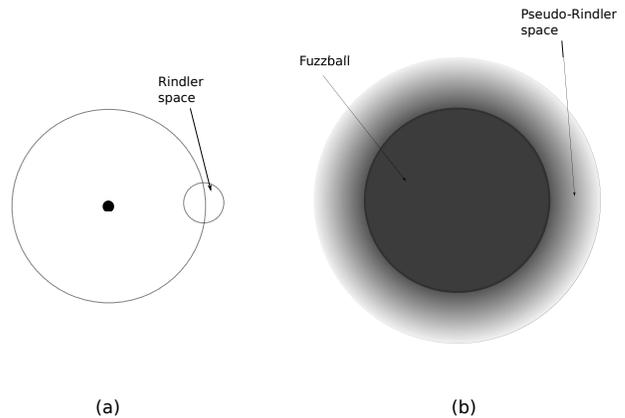}
\caption{\label{figrindlercausality} (a) In the traditional hole, the region just {\it outside} the horizon is Rindler space, which is just a part of Minkowski space. (b) The dark circle is the fuzzball. The region outside the fuzzball has extra  vacuum fluctuations that correspond to the fuzzball of mass $M$ fluctuating to a fuzzball of mass $M_f>M$. These virtual fuzzballs are depicted by the shaded region outside the fuzzball. Because of altered vacuum fluctuations, the region near the fuzzball boundary is  termed pseudo-Rindler space.}
\end{figure}

Let us now summarize our central proposal:

\mm

(i) In fig.\ref{fig1p}(a) we depict the traditional picture of the black hole, where we have the vacuum state around the horizon $r=2GM$. In fig.\ref{fig1p}(b) we depict the fuzzball, which has  a boundary at a location 
\be
r\approx 2GM+\epsilon\equiv r_b
\ee
where $\epsilon \ll GM$. We wish to argue that vacuum fluctuations in the region $r>r_b$ in the fuzzball spacetime are different from the vacuum fluctuations outside the horizon in the traditional black hole. 

\mm

(ii) The dashed lines in fig.\ref{fig1p}(b) depict the processes that modify the vacuum outside the fuzzball.  We conjecture that the fuzzball of mass $M$ undergoes virtual fluctuations to fuzzballs of mass $M_f>M$; the dashed lines therefore represent processes that create such fuzzballs.  The energy $M_f-M$ is not small, so one might think that such fluctuations would be suppressed. But we conjecture that these fluctuations nevertheless have a nontrivial effect because of  `entropy enhancement': there are a large  number of possible fuzzballs  of mass $M_f$ that the system can fluctuate to. 

\mm

(iii)    This altered vacuum polarization has very little effect on low energy infalling objects, so these objects just see the normal Schwarzschild metric for a black hole of mass $M$ in the region  $r>r_b$. But an infalling object with high energy ($E\gg T$) converts the virtual fluctuations of the fuzzball  to on-shell fuzzball states before it reaches $r=r_b$; the extra energy required for the larger fuzzball is drawn from the energy of the infalling object.  This effect prevents the infalling object from getting trapped inside its own horizon: energy is leaked away to fuzzballs just before a horizon would have formed.  The space near the fuzzball boundary  $r=r_b$ is termed `pseudo-Rindler space' to emphasize the fact that it has a different vacuum polarization from usual Rindler space which is just a part of normal  Minkowski space (fig.\ref{figrindlercausality}).

\mm

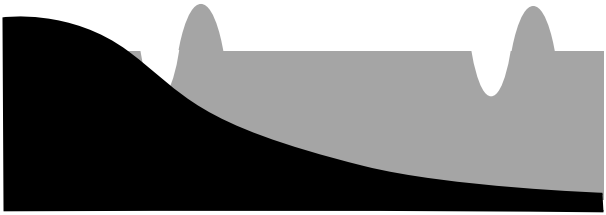
\begin{figure}

\definecolor{ca5a5a5}{RGB}{165,165,165}
\definecolor{cffffff}{RGB}{255,255,255}

\begin{tikzpicture}[y=0.80pt, x=0.80pt, yscale=-.17000000, xscale=.17000000, inner sep=0pt, outer sep=0pt]
  \path[fill=ca5a5a5,line join=miter,line cap=butt,miter limit=4.00,even odd
    rule,line width=0.000pt] (-708.5714,260.9336) .. controls (-708.5714,260.9336)
    and (-687.9726,126.6479) .. (-646.7750,126.6479) .. controls
    (-605.5773,126.6479) and (-585.7142,258.5357) .. (-585.7142,258.5357) --
    cycle;
  \path[fill=ca5a5a5,miter limit=4.00,draw opacity=0.986,line
    width=0.000pt,rounded corners=0.0000cm] (-1831.4286,252.3622) rectangle
    (-451.4286,666.6479);
  \path[fill=ca5a5a5,line join=miter,line cap=butt,miter limit=4.00,even odd
    rule,line width=0.000pt] (-1634.2857,255.2193) .. controls
    (-1634.2857,255.2193) and (-1613.2078,120.9336) .. (-1571.0522,120.9336) ..
    controls (-1528.8964,120.9336) and (-1508.5713,252.8214) ..
    (-1508.5713,252.8214) -- cycle;
  \path[fill=cffffff,line join=miter,line cap=butt,miter limit=4.00,even odd
    rule,line width=0.000pt] (-820.0000,243.7908) .. controls (-820.0000,243.7908)
    and (-801.3173,378.0765) .. (-763.9521,378.0765) .. controls
    (-726.5867,378.0765) and (-708.5713,246.1887) .. (-708.5713,246.1887) --
    cycle;
  \path[fill=cffffff,line join=miter,line cap=butt,miter limit=4.00,even odd
    rule,line width=0.000pt] (-1739.9999,246.6479) .. controls
    (-1739.9999,246.6479) and (-1721.7963,380.9336) .. (-1685.3891,380.9336) ..
    controls (-1648.9819,380.9336) and (-1631.4284,249.0458) ..
    (-1631.4284,249.0458) -- cycle;
  \path[draw=black,fill=black,line join=miter,line cap=butt,even odd rule,line
    width=0.800pt] (-451.4286,698.0765) .. controls (-1000.0000,692.3622) and
    (-1567.1282,693.3293) .. (-2117.1428,695.2193) -- (-2120.0000,160.9336) ..
    controls (-2120.0000,160.9336) and (-1934.2857,135.2193) ..
    (-1774.2857,255.2193) .. controls (-1614.2857,375.2193) and
    (-1597.1428,455.2193) .. (-1117.1428,575.2193) .. controls
    (-881.0970,634.2308) and (-454.2857,649.5050) .. (-454.2857,649.5050);

\end{tikzpicture}

\caption{  The black region is land, while the grey region is water.  A wave of  travels freely when its amplitude is much less than the depth of the water, but will suffer nontrivial deformation when the amplitude becomes comparable to the depth of the water.}\label{fig3}

\end{figure}

We can get a schematic model of the above conjecture by considering the lake depicted in fig.\ref{fig3}.  On the left is land; this represents the interior of the fuzzball $r < r_b$. The water represents the exterior region $r>r_b$. Waves can propagate on the surface of this water, and represent matter quanta in the region $r>r_b$. The vacuum fluctuations are strongest near the fuzzball boundary, and their effect is to reduce the depth of the lake to a  small value near the fuzzball surface. Low energy waves propagate without noticing the reduced depth, all the way upto $r=\approx r_b$. But large waves (representing high energy infalling objects) feel the finite depth of the lake at some location $r>r_b$, and their evolution  changes at this point. In our actual problem this `bottom of the lake' is felt when the infalling object carrying energy $\Delta M$ reaches $r\approx 2G(M+\Delta M)$; the semiclassical evolution then becomes invalid and an `entropy enhanced tunneling' takes place. As a consequence, the objet is never trapped in its own horizon, and we avoid any problem with causality. 

We proceed in the following steps:

\mm

(a) We describe a  toy model for the fuzzball; this will help explain how the no-hair theorems and the Buchdahl limit are bypassed by the fuzzball structure found in string theory (section \ref{sec:2}).

\mm

(b) We use this toy model to give a picture of gravitational collapse: an infalling shell starts nucleating  fuzzball excitations   as it approaches the location $r\approx 2GM$, and thereby ends up as a fuzzball of radius $r_b$ rather than a spacetime with horizon (section \ref{sec:3}).  

\mm

(c) We use some toy models and analogies to explain our proposal that the vacuum fluctuations outside the fuzzball are different from the fluctuations of empty space, and how such a change can lead to fuzzball formation before an infalling object reaches $r=r_b$ (section \ref{sec:6}). 

\mm

(d) We state our proposal in concrete form, and explain how it allows us to preserve causality in the process of black hole formation and evaporation (sections \ref{sec:7} and \ref{sec:8c}). 

\mm

(e) We give some rough estimates of the location where the `entropy enhanced tunneling' is expected to take place in different instances of infall (section \ref{sec:8}). 

\mm 

(f) We note the difference between our picture of fuzzball dynamics and the assumed dynamics in the firewall argument \cite{amps}; we argue that the assumptions in the firewall argument are in conflict with each other because of a violation of causality (section \ref{sec:4}). 

\mm

(g) We recall the conjecture of fuzzball complementarity, and note that causality in the underlying theory  is a logical requirement for any such conjecture (section \ref{sec:4c}). 

\mm

(h) We close with a summary and a general discussion of causality (section \ref{sec:10}).

\section{\label{sec:2}A toy model for the fuzzball}

Fuzzballs are solutions found in the full quantum theory of strings. But  we can understand some essential aspects of their structure by looking at  toy models found in  Einstein gravity with an extra dimension. In this section we recall one  such a solution described in \cite{model};  one may use this toy model as a rough picture to understand the elements that compose an actual fuzzball solution. We first mention the problem that fuzzballs solve, and then give the model for the fuzzball. 

A key aspect of the information paradox  is the argument that  `black holes have no hair'. Consider the Schwarzschild metric
\be
ds^2=-(1-{2M\over r}) dt^2 +{dr^2\over 1-{2M\over r}}+r^2d\Omega_2^2
\label{qtwo}
\ee
We can consider a scalar $\square \phi=0$ and try to add scalar `hair' of the form
\be
\phi=Re[\t \phi(r) Y_{lm}(\theta, \phi) e^{-i\omega t}]
\ee
But we find that there are no regular solutions for the function $\t \phi(r)$: rapid oscillations near the horizon lead to a divergence of the  stress tensor of $\phi$ at the location  $r\r 2M$. This is an illustration of the classical `no hair' theorem, but it is crucial that the same computation leads gives `no hair' at the {\it quantum} level. To see this, we first recall a situation where we {\it can} add hair to deform the solution. Consider a static star given by the metric
\be
ds^2=-f(r) dt^2 +{dr^2 \over g(r)}+r^2d\Omega_2^2
\label{qone}
\ee
Let the lowest energy state for the quantum field $\hat\phi$ in this vacuum be $|0\rangle_{star}$. We can again solve the wave equation $\square \phi=0$ in the metric (\ref{qone}) 
\be
\phi=\t \phi_{klm}(r) Y_{lm}(\theta, \phi) e^{-i\omega_{kl} t}\equiv \phi_{klm}
\ee
This time there will be a complete set of solutions of this form. We can therefore write
\be
\hat \phi = \sum_{klm} \left ( \phi_{klm} \hat a_{klm} + \phi^*_{klm} \hat a^\dagger_{klm}\right )
\ee
Then the states of the system are given by exciting the vacuum $|0\rangle_{star}$ 
\be
\hat a^\dagger_{k_1l_1m_1} \dots \hat a^\dagger_{k_n l_n m_n} |0\rangle_{star}
\label{qtfive}
\ee
These excitations add `scalar hair' to the star.  

By contrast,  for the black hole metric (\ref{qtwo}) we do not find regular solutions $\phi_{klm}$, and so we cannot change the quantum state in this way. The quantum vacuum state  around the horizon is therefore unique, and it is this state that leads to the creation of Hawking's entangled pairs. 

We have looked at a simple example above, but years of effort with different models did not shake this basic conclusion that the horizon cannot be deformed, either at the classical or quantum level.  One may try to avoid the problem by not allowing a horizon in the first place, but here we run into results like Buchdahl's theorem \cite{buchdahl}. Consider a star made of a perfect fluid, whose density $\rho$ increases monotonically inwards. If the radius $R$ of the fluid ball satisfies
\be
R<{9M\over 4}
\label{qthree}
\ee
then the pressure will diverge at some radius $r>0$, rendering the solution invalid. Thus any fluid ball that has been compressed to a size smaller than (\ref{qthree}) must necessarily collapse and generate a horizon. 

String theory avoids these problems in a remarkable way, through the {\it fuzzball mechanism}. A toy model for the fuzzball was discussed in \cite{model}. Consider the 4+1 dimensional spacetime obtained by adding a trivial time direction to the 3+1 dimensional Euclidean Schwarzschild solution
\be
ds^2=-dt^2 + (1-{r_0\over r})d\tau^2 + {dr^2\over 1-{r_0\over r}} + r^2 (d\theta^2+\sin^2\theta d\phi^2)
\label{metrickk}
\ee
This metric is a perfectly regular solution of the 4+1 vacuum Einstein equations. The `Euclidean time' direction $\tau$ is compact, with $0\le \tau < 4\pi r_0$. The $r_0, \tau$ directions form a cigar, whose tip lies at $r=r_0$.  The spacetime ends at $r=r_0$; we can say that the ball $r<r_0$ has been excised from the manifold, and the compact directions closed off to generate a geodesically complete spacetime.  

We can now dimensionally reduce on the circle $\tau$, regarding this solution as a 3+1 dimensional metric in $(t, r, \theta, \phi)$ coupled to a scalar field 
\be
\Phi={\sqrt{3}\over 2}\ln  (1-{r_0\over r})
\ee
describing the radius of the compact direction $\tau$. This scalar field has a standard stress tensor, whose value works out to be
\be
T^\mu{}_{\nu} =  {\rm diag }\{-\rho, p_r, p_\theta, p_\phi    \}={\rm diag }\{-f, f, -f, -f    \} 
\ee
where
\be
f= {3r_0^2\over 8r^4 (1-{r_0\over r})^{3\over 2}}
\ee
We see that the pressures do diverge at $r\r r_0>0$, and if we followed the spirit of Buchdahl's theorem, we would discard this solution. But the solution is actually  a perfectly regular solution in 4+1 dimensions; what breaks down is the dimensional reduction map when the length of the compact circle goes to zero. 

The fuzzball solutions are similar in spirit: they are valid solutions in the full 10-dimensional string theory, but are singular when viewed from the perspective of the noncompact directions alone. The simplest fuzzball solutions are characterized by a set of KK monopoles and antimonopoles, which are regular solutions of gravity but with a singular dimensional reduction.  In the limit where the center if a KK monopole coincides with the center of an anti-monopole, it is known that the bosonic fields yield the Euclidean Schwarzschild solution tensored trivially with time and the other compact directions, similar in spirit to (\ref{metrickk}) \cite{sen}. This solution is unstable, but that is as it should be:  microstates of the nonextremal hole should radiate radiate energy, and in specific cases the instability of the microstate solutions has been shown to map exactly to the Hawking radiation expected from that microstate \cite{radiation}. 

Thus we can use the solution (\ref{metrickk}) as our toy model of a fuzzball to illustrate the picture of the gravitational vacuum that we wish to present. 

Before proceeding, we note that  similar features are obtained for the `bubble of nothing' which was discovered an an instability of the vacuum for the spacetime $M_{3,1} \times S^1$ \cite{witten}. In this solution a bubble nucleates by a vacuum fluctuations that pinches off the compact circle. After the bubble tunnels to a certain size, it can continue to expand further as an on-shell classical solution.   Topologically the metric is similar to the Euclidean Schwarzschild solution. Again we can dimensionally reduce on the $S^1$, getting a scalar $\Phi$ on 3+1 dimensional spacetime.  The stress tensor of $\Phi$  diverges as we approach the bubble wall, though the overall spacetime is smooth \cite{beyond,model}. In our qualitative analysis below we will use the Euclidean Schwarzschild solution as our toy model of the fuzzball, and we will assume (by analogy with the bubble of nothing) that such solutions can nucleate by tunneling when a suitable amount of energy is available.  

The actual fuzzball solutions involve other fields of string theory besides the metric. We may roughly picture such a fuzzball as having many KK monopoles and antimonopoles, with fluxes on the spheres between these topological objects. It was explained in \cite{gibbonswarner} how such solutions evade the conditions assumed in deriving the various types of no-hair results in earlier years.  

\section{\label{sec:3}Modelling the tunneling into fuzzballs}

As noted in section \ref{sec:1}, the fuzzball paradigm says that a collapsing shell suffers an `entropy enhanced tunneling' to fuzzballs. In this section we conjecture a picture of when and where this tunneling should take place; this picture will be consistent with causality of the underlying gravity theory. In later sections we will argue that such a  picture is  made possible by an altered polarization of the vacuum outside the fuzzball (the change from Rindler to `pseudo-Rindler'). 

Let the theory of gravity be such that all black hole microstates are fuzzballs.  To picture these fuzzballs,  we can imagine that spacetime has the topology $M_{3,1}\times S^1$, and that the radius of $S^1$ is $4\pi r_0$. Then around any point of space we can nucleate a `bubble' of the form (\ref{metrickk}), where a sphere of radius $r_0$ has been removed and the compact direction smoothly closed off at the boundary of this sphere. A general fuzzball state in string theory may be pictured as having many such bubbles with other objects like flux-carrying spheres carrying spheres linking the bubbles.

\begin{figure}

\begin{tikzpicture}[y=0.80pt, x=0.80pt, yscale=-.09000000, xscale=.09000000, inner sep=0pt, outer sep=0pt]
  \begin{scope}[shift={(-471.42857,28.57143)}]
    \path[draw=black,dash pattern=on 1.97pt off 1.97pt,line join=miter,line
      cap=butt,miter limit=4.00,even odd rule,line width=0.986pt]
      (-971.4286,-190.4949) ellipse (13.2611cm and 13.1805cm);
    \path[draw=black,line join=miter,line cap=butt,miter limit=4.00,even odd
      rule,line width=3.098pt] (-971.4286,-190.4949) ellipse (17.4048cm and
      17.5661cm);
  \end{scope}
  \path[fill=black,line join=miter,line cap=butt,line width=0.800pt]
    (-1491.7170,682.8970) node[above right] (text4470) {(a)};
  \path[fill=black,line join=miter,line cap=butt,line width=0.800pt]
    (317.5994,682.9312) node[above right] (text4474) {(b)};
    \path[draw=black,dash pattern=on 1.45pt off 1.45pt,line join=miter,line
      cap=butt,miter limit=4.00,even odd rule,line width=0.725pt]
      (375.7143,-154.7806) circle (9.7178cm);
    \path[draw=black,line join=miter,line cap=butt,miter limit=4.00,even odd
      rule,line width=1.888pt] (375.7143,-154.7806) circle (12.4615cm);
    \path[draw=black,line join=miter,line cap=butt,miter limit=4.00,even odd
      rule,line width=0.640pt] (254.2856,-679.0664) circle (0.9031cm);
    \path[cm={{0.70711,-0.70711,0.70711,0.70711,(0.0,0.0)}},draw=black,line
      join=miter,line cap=butt,miter limit=4.00,even odd rule,line width=0.480pt]
      (556.9441,-302.3856) ellipse (2.0320cm and 0.5080cm);
    \path[cm={{0.97557,-0.21967,0.14792,0.989,(0.0,0.0)}},draw=black,line
      join=miter,line cap=butt,miter limit=4.00,even odd rule,line width=0.414pt]
      (98.2324,-503.0207) ellipse (1.5210cm and 0.5039cm);
    \path[cm={{-0.05183,-0.99866,0.98513,-0.17183,(0.0,0.0)}},draw=black,line
      join=miter,line cap=butt,miter limit=4.00,even odd rule,line width=0.356pt]
      (433.3398,-38.1080) ellipse (1.1146cm and 0.5086cm);
    \path[cm={{-0.61729,-0.78673,0.86996,-0.49312,(0.0,0.0)}},draw=black,line
      join=miter,line cap=butt,miter limit=4.00,even odd rule,line width=0.401pt]
      (206.1191,-29.4514) ellipse (1.6452cm and 0.4376cm);
    \path[draw=black,line join=miter,line cap=butt,miter limit=4.00,even odd
      rule,line width=0.640pt] (105.7143,-539.0664) circle (0.9031cm);
    \path[draw=black,line join=miter,line cap=butt,miter limit=4.00,even odd
      rule,line width=0.640pt] (-62.8572,-499.0664) circle (0.9031cm);
    \path[draw=black,line join=miter,line cap=butt,miter limit=4.00,even odd
      rule,line width=0.640pt] (-202.8572,-210.4950) circle (0.9031cm);
    \path[draw=black,line join=miter,line cap=butt,miter limit=4.00,even odd
      rule,line width=0.640pt] (-57.1429,-356.2093) circle (0.9031cm);
    \path[cm={{0.70711,-0.70711,0.70711,0.70711,(0.0,0.0)}},draw=black,line
      join=miter,line cap=butt,miter limit=4.00,even odd rule,line width=0.480pt]
      (112.4770,-290.2637) ellipse (2.0320cm and 0.5080cm);
    \path[cm={{-0.66187,-0.74962,0.79591,-0.60542,(0.0,0.0)}},draw=black,line
      join=miter,line cap=butt,miter limit=4.00,even odd rule,line width=0.414pt]
      (-70.8148,-107.3516) ellipse (1.5210cm and 0.5039cm);
    \path[cm={{0.05183,0.99866,-0.98513,0.17183,(0.0,0.0)}},draw=black,line
      join=miter,line cap=butt,miter limit=4.00,even odd rule,line width=0.356pt]
      (-28.6468,97.1022) ellipse (1.1146cm and 0.5086cm);
    \path[scale=-1.000,draw=black,line join=miter,line cap=butt,miter
      limit=4.00,even odd rule,line width=0.640pt] (-37.1428,-178.0765) circle
      (0.9031cm);
    \path[scale=-1.000,draw=black,line join=miter,line cap=butt,miter
      limit=4.00,even odd rule,line width=0.640pt] (94.2856,-60.9338) circle
      (0.9031cm);
    \path[scale=-1.000,draw=black,line join=miter,line cap=butt,miter
      limit=4.00,even odd rule,line width=0.640pt] (100.0000,81.9235) circle
      (0.9031cm);
    \path[draw=black,line join=miter,line cap=butt,miter limit=4.00,even odd
      rule,line width=0.640pt] (99.9999,369.5050) circle (0.9031cm);
    \path[cm={{-0.23861,-0.97112,0.97112,-0.23861,(0.0,0.0)}},draw=black,line
      join=miter,line cap=butt,miter limit=4.00,even odd rule,line width=0.480pt]
      (-280.8569,1.6033) ellipse (2.0320cm and 0.5080cm);
    \path[cm={{0.63996,-0.7684,0.4865,0.87368,(0.0,0.0)}},draw=black,line
      join=miter,line cap=butt,miter limit=4.00,even odd rule,line width=0.228pt]
      (-35.5051,334.4735) ellipse (0.7970cm and 0.2911cm);
    \path[cm={{-0.7825,-0.62265,0.87864,-0.47748,(0.0,0.0)}},draw=black,line
      join=miter,line cap=butt,miter limit=4.00,even odd rule,line width=0.257pt]
      (470.1735,773.1666) ellipse (0.9230cm and 0.3201cm);
    \path[draw=black,line join=miter,line cap=butt,miter limit=4.00,even odd
      rule,line width=0.640pt] (182.8571,278.0766) circle (0.9031cm);
    \path[draw=black,line join=miter,line cap=butt,miter limit=4.00,even odd
      rule,line width=0.640pt] (377.1429,-633.3521) circle (0.9031cm);
    \path[draw=black,line join=miter,line cap=butt,miter limit=4.00,even odd
      rule,line width=0.640pt] (368.5715,318.0765) circle (0.9031cm);
    \path[cm={{0.89985,0.43621,-0.43621,0.89985,(0.0,0.0)}},draw=black,line
      join=miter,line cap=butt,miter limit=4.00,even odd rule,line width=0.480pt]
      (386.9658,156.3681) ellipse (2.0320cm and 0.5080cm);
    \path[scale=-1.000,draw=black,line join=miter,line cap=butt,miter
      limit=4.00,even odd rule,line width=0.640pt] (-937.1428,101.9235) circle
      (0.9031cm);
    \path[scale=-1.000,draw=black,line join=miter,line cap=butt,miter
      limit=4.00,even odd rule,line width=0.640pt] (-489.4067,-366.6479) circle
      (0.9031cm);
    \path[cm={{-0.70711,0.70711,-0.70711,-0.70711,(0.0,0.0)}},draw=black,line
      join=miter,line cap=butt,miter limit=4.00,even odd rule,line width=0.480pt]
      (-189.8390,-607.3423) ellipse (2.0320cm and 0.5080cm);
    \path[cm={{-0.97557,0.21967,-0.14792,-0.989,(0.0,0.0)}},draw=black,line
      join=miter,line cap=butt,miter limit=4.00,even odd rule,line width=0.414pt]
      (-685.5795,-361.2249) ellipse (1.5210cm and 0.5039cm);
    \path[cm={{0.05183,0.99866,-0.98513,0.17183,(0.0,0.0)}},draw=black,line
      join=miter,line cap=butt,miter limit=4.00,even odd rule,line width=0.356pt]
      (252.0342,-802.5671) ellipse (1.1146cm and 0.5086cm);
    \path[cm={{0.61676,0.78715,-0.86966,0.49365,(0.0,0.0)}},draw=black,line
      join=miter,line cap=butt,miter limit=4.00,even odd rule,line width=0.327pt]
      (291.4953,-815.9766) ellipse (1.3428cm and 0.3570cm);
    \path[scale=-1.000,draw=black,line join=miter,line cap=butt,miter
      limit=4.00,even odd rule,line width=0.640pt] (-637.9783,-236.6478) circle
      (0.9031cm);
    \path[scale=-1.000,draw=black,line join=miter,line cap=butt,miter
      limit=4.00,even odd rule,line width=0.640pt] (-806.5496,-186.6479) circle
      (0.9031cm);
    \path[scale=-1.000,draw=black,line join=miter,line cap=butt,miter
      limit=4.00,even odd rule,line width=0.640pt] (-800.8352,-43.7908) circle
      (0.9031cm);
    \path[cm={{-0.70711,0.70711,-0.70711,-0.70711,(0.0,0.0)}},draw=black,line
      join=miter,line cap=butt,miter limit=4.00,even odd rule,line width=0.480pt]
      (-634.3062,-595.2205) ellipse (2.0320cm and 0.5080cm);
    \path[cm={{0.66187,0.74962,-0.79591,0.60542,(0.0,0.0)}},draw=black,line
      join=miter,line cap=butt,miter limit=4.00,even odd rule,line width=0.414pt]
      (131.3153,-873.6587) ellipse (1.5210cm and 0.5039cm);
    \path[cm={{-0.05183,-0.99866,0.98513,-0.17183,(0.0,0.0)}},draw=black,line
      join=miter,line cap=butt,miter limit=4.00,even odd rule,line width=0.356pt]
      (152.6587,861.5613) ellipse (1.1146cm and 0.5086cm);
    \path[draw=black,line join=miter,line cap=butt,miter limit=4.00,even odd
      rule,line width=0.640pt] (706.5496,-490.4950) circle (0.9031cm);
    \path[draw=black,line join=miter,line cap=butt,miter limit=4.00,even odd
      rule,line width=0.640pt] (837.9780,-373.3523) circle (0.9031cm);
    \path[draw=black,line join=miter,line cap=butt,miter limit=4.00,even odd
      rule,line width=0.640pt] (843.6924,-230.4950) circle (0.9031cm);
    \path[scale=-1.000,draw=black,line join=miter,line cap=butt,miter
      limit=4.00,even odd rule,line width=0.640pt] (-643.6924,681.9236) circle
      (0.9031cm);
    \path[cm={{0.23861,0.97112,-0.97112,0.23861,(0.0,0.0)}},draw=black,line
      join=miter,line cap=butt,miter limit=4.00,even odd rule,line width=0.480pt]
      (-406.8013,-795.1537) ellipse (2.0320cm and 0.5080cm);
    \path[cm={{-0.63996,0.7684,-0.4865,-0.87368,(0.0,0.0)}},draw=black,line
      join=miter,line cap=butt,miter limit=4.00,even odd rule,line width=0.228pt]
      (-894.8624,-63.7450) ellipse (0.7970cm and 0.2911cm);
    \path[cm={{0.7825,0.62265,-0.87864,0.47748,(0.0,0.0)}},draw=black,line
      join=miter,line cap=butt,miter limit=4.00,even odd rule,line width=0.257pt]
      (557.7043,4.7106) ellipse (0.9230cm and 0.3201cm);
    \path[scale=-1.000,draw=black,line join=miter,line cap=butt,miter
      limit=4.00,even odd rule,line width=0.640pt] (-566.5496,585.6380) circle
      (0.9031cm);
    \path[cm={{-0.89985,-0.43621,0.43621,-0.89985,(0.0,0.0)}},draw=black,line
      join=miter,line cap=butt,miter limit=4.00,even odd rule,line width=0.480pt]
      (-145.9623,761.9019) ellipse (2.0320cm and 0.5080cm);
  \begin{scope}[shift={(0,8.57143)}]
    \path[fill=black,line join=miter,line cap=butt,line width=0.800pt]
      (-1462.4006,2505.7883) node[above right] (text4474-6) {(c)};
      \path[draw=black,dash pattern=on 0.86pt off 0.86pt,line join=miter,line
        cap=butt,miter limit=4.00,even odd rule,line width=0.428pt]
        (-1402.8572,1686.6479) circle (5.7316cm);
      \path[draw=black,line join=miter,line cap=butt,miter limit=4.00,even odd
        rule,line width=1.126pt] (-1402.8572,1686.6479) circle (7.4352cm);
      \path[draw=black,line join=miter,line cap=butt,miter limit=4.00,even odd
        rule,line width=0.640pt] (-1525.7144,1143.7906) circle (0.9031cm);
      \path[cm={{0.70711,-0.70711,0.70711,0.70711,(0.0,0.0)}},draw=black,line
        join=miter,line cap=butt,miter limit=4.00,even odd rule,line width=0.480pt]
        (-1990.6606,-272.0811) ellipse (2.0320cm and 0.5080cm);
      \path[cm={{0.97557,-0.21967,0.14792,0.989,(0.0,0.0)}},draw=black,line
        join=miter,line cap=butt,miter limit=4.00,even odd rule,line width=0.414pt]
        (-1937.2399,887.9988) ellipse (1.5210cm and 0.5039cm);
      \path[cm={{-0.05183,-0.99866,0.98513,-0.17183,(0.0,0.0)}},draw=black,line
        join=miter,line cap=butt,miter limit=4.00,even odd rule,line width=0.356pt]
        (-1067.4927,-1923.9407) ellipse (1.1146cm and 0.5086cm);
      \path[cm={{-0.61729,-0.78673,0.86996,-0.49312,(0.0,0.0)}},draw=black,line
        join=miter,line cap=butt,miter limit=4.00,even odd rule,line width=0.401pt]
        (-509.9290,-2583.6077) ellipse (1.6452cm and 0.4376cm);
      \path[draw=black,line join=miter,line cap=butt,miter limit=4.00,even odd
        rule,line width=0.640pt] (-1674.2858,1283.7908) circle (0.9031cm);
      \path[draw=black,line join=miter,line cap=butt,miter limit=4.00,even odd
        rule,line width=0.640pt] (-1842.8572,1323.7906) circle (0.9031cm);
      \path[draw=black,line join=miter,line cap=butt,miter limit=4.00,even odd
        rule,line width=0.640pt] (-1982.8572,1612.3621) circle (0.9031cm);
      \path[draw=black,line join=miter,line cap=butt,miter limit=4.00,even odd
        rule,line width=0.640pt] (-1837.1429,1466.6478) circle (0.9031cm);
      \path[cm={{0.70711,-0.70711,0.70711,0.70711,(0.0,0.0)}},draw=black,line
        join=miter,line cap=butt,miter limit=4.00,even odd rule,line width=0.480pt]
        (-2435.1277,-259.9591) ellipse (2.0320cm and 0.5080cm);
      \path[cm={{-0.66187,-0.74962,0.79591,-0.60542,(0.0,0.0)}},draw=black,line
        join=miter,line cap=butt,miter limit=4.00,even odd rule,line width=0.414pt]
        (-444.9843,-2654.9565) ellipse (1.5210cm and 0.5039cm);
      \path[cm={{0.05183,0.99866,-0.98513,0.17183,(0.0,0.0)}},draw=black,line
        join=miter,line cap=butt,miter limit=4.00,even odd rule,line width=0.356pt]
        (1472.1857,1982.9349) ellipse (1.1146cm and 0.5086cm);
      \path[scale=-1.000,draw=black,line join=miter,line cap=butt,miter
        limit=4.00,even odd rule,line width=0.640pt] (1742.8572,-2000.9336) circle
        (0.9031cm);
      \path[scale=-1.000,draw=black,line join=miter,line cap=butt,miter
        limit=4.00,even odd rule,line width=0.640pt] (1874.2856,-1883.7909) circle
        (0.9031cm);
      \path[scale=-1.000,draw=black,line join=miter,line cap=butt,miter
        limit=4.00,even odd rule,line width=0.640pt] (1880.0000,-1740.9336) circle
        (0.9031cm);
      \path[draw=black,line join=miter,line cap=butt,miter limit=4.00,even odd
        rule,line width=0.640pt] (-1680.0001,2192.3621) circle (0.9031cm);
      \path[cm={{-0.23861,-0.97112,0.97112,-0.23861,(0.0,0.0)}},draw=black,line
        join=miter,line cap=butt,miter limit=4.00,even odd rule,line width=0.480pt]
        (-1626.3423,-2161.9302) ellipse (2.0320cm and 0.5080cm);
      \path[cm={{0.63996,-0.7684,0.4865,0.87368,(0.0,0.0)}},draw=black,line
        join=miter,line cap=butt,miter limit=4.00,even odd rule,line width=0.228pt]
        (-2652.9680,118.8193) ellipse (0.7970cm and 0.2911cm);
      \path[cm={{-0.7825,-0.62265,0.87864,-0.47748,(0.0,0.0)}},draw=black,line
        join=miter,line cap=butt,miter limit=4.00,even odd rule,line width=0.257pt]
        (-346.2951,-1979.8077) ellipse (0.9230cm and 0.3201cm);
      \path[draw=black,line join=miter,line cap=butt,miter limit=4.00,even odd
        rule,line width=0.640pt] (-1597.1429,2100.9336) circle (0.9031cm);
      \path[draw=black,line join=miter,line cap=butt,miter limit=4.00,even odd
        rule,line width=0.640pt] (-1402.8572,1189.5050) circle (0.9031cm);
      \path[draw=black,line join=miter,line cap=butt,miter limit=4.00,even odd
        rule,line width=0.640pt] (-1411.4285,2140.9336) circle (0.9031cm);
      \path[cm={{0.89985,0.43621,-0.43621,0.89985,(0.0,0.0)}},draw=black,line
        join=miter,line cap=butt,miter limit=4.00,even odd rule,line width=0.480pt]
        (-419.6118,2573.1099) ellipse (2.0320cm and 0.5080cm);
      \path[scale=-1.000,draw=black,line join=miter,line cap=butt,miter
        limit=4.00,even odd rule,line width=0.640pt] (842.8572,-1720.9336) circle
        (0.9031cm);
      \path[scale=-1.000,draw=black,line join=miter,line cap=butt,miter
        limit=4.00,even odd rule,line width=0.640pt] (1290.5933,-2189.5051) circle
        (0.9031cm);
      \path[cm={{-0.70711,0.70711,-0.70711,-0.70711,(0.0,0.0)}},draw=black,line
        join=miter,line cap=butt,miter limit=4.00,even odd rule,line width=0.480pt]
        (2357.7656,-637.6468) ellipse (2.0320cm and 0.5080cm);
      \path[cm={{-0.97557,0.21967,-0.14792,-0.989,(0.0,0.0)}},draw=black,line
        join=miter,line cap=butt,miter limit=4.00,even odd rule,line width=0.414pt]
        (1349.8927,-1752.2445) ellipse (1.5210cm and 0.5039cm);
      \path[cm={{0.05183,0.99866,-0.98513,0.17183,(0.0,0.0)}},draw=black,line
        join=miter,line cap=butt,miter limit=4.00,even odd rule,line width=0.356pt]
        (1752.8667,1083.2656) ellipse (1.1146cm and 0.5086cm);
      \path[cm={{0.61676,0.78715,-0.86966,0.49365,(0.0,0.0)}},draw=black,line
        join=miter,line cap=butt,miter limit=4.00,even odd rule,line width=0.327pt]
        (1005.9234,1737.4598) ellipse (1.3428cm and 0.3570cm);
      \path[scale=-1.000,draw=black,line join=miter,line cap=butt,miter
        limit=4.00,even odd rule,line width=0.640pt] (1142.0217,-2059.5049) circle
        (0.9031cm);
      \path[scale=-1.000,draw=black,line join=miter,line cap=butt,miter
        limit=4.00,even odd rule,line width=0.640pt] (973.4504,-2009.5050) circle
        (0.9031cm);
      \path[scale=-1.000,draw=black,line join=miter,line cap=butt,miter
        limit=4.00,even odd rule,line width=0.640pt] (979.1648,-1866.6478) circle
        (0.9031cm);
      \path[cm={{-0.70711,0.70711,-0.70711,-0.70711,(0.0,0.0)}},draw=black,line
        join=miter,line cap=butt,miter limit=4.00,even odd rule,line width=0.480pt]
        (1913.2985,-625.5250) ellipse (2.0320cm and 0.5080cm);
      \path[cm={{0.66187,0.74962,-0.79591,0.60542,(0.0,0.0)}},draw=black,line
        join=miter,line cap=butt,miter limit=4.00,even odd rule,line width=0.414pt]
        (505.4849,1673.9463) ellipse (1.5210cm and 0.5039cm);
      \path[cm={{-0.05183,-0.99866,0.98513,-0.17183,(0.0,0.0)}},draw=black,line
        join=miter,line cap=butt,miter limit=4.00,even odd rule,line width=0.356pt]
        (-1348.1738,-1024.2715) ellipse (1.1146cm and 0.5086cm);
      \path[draw=black,line join=miter,line cap=butt,miter limit=4.00,even odd
        rule,line width=0.640pt] (-1073.4504,1332.3621) circle (0.9031cm);
      \path[draw=black,line join=miter,line cap=butt,miter limit=4.00,even odd
        rule,line width=0.640pt] (-942.0220,1449.5048) circle (0.9031cm);
      \path[draw=black,line join=miter,line cap=butt,miter limit=4.00,even odd
        rule,line width=0.640pt] (-936.3076,1592.3621) circle (0.9031cm);
      \path[scale=-1.000,draw=black,line join=miter,line cap=butt,miter
        limit=4.00,even odd rule,line width=0.640pt] (1136.3076,-1140.9335) circle
        (0.9031cm);
      \path[cm={{0.23861,0.97112,-0.97112,0.23861,(0.0,0.0)}},draw=black,line
        join=miter,line cap=butt,miter limit=4.00,even odd rule,line width=0.480pt]
        (938.6842,1368.3796) ellipse (2.0320cm and 0.5080cm);
      \path[cm={{-0.63996,0.7684,-0.4865,-0.87368,(0.0,0.0)}},draw=black,line
        join=miter,line cap=butt,miter limit=4.00,even odd rule,line width=0.228pt]
        (1722.6005,151.9091) ellipse (0.7970cm and 0.2911cm);
      \path[cm={{0.7825,0.62265,-0.87864,0.47748,(0.0,0.0)}},draw=black,line
        join=miter,line cap=butt,miter limit=4.00,even odd rule,line width=0.257pt]
        (1374.1729,2757.6851) ellipse (0.9230cm and 0.3201cm);
      \path[scale=-1.000,draw=black,line join=miter,line cap=butt,miter
        limit=4.00,even odd rule,line width=0.640pt] (1213.4504,-1237.2191) circle
        (0.9031cm);
      \path[cm={{-0.89985,-0.43621,0.43621,-0.89985,(0.0,0.0)}},draw=black,line
        join=miter,line cap=butt,miter limit=4.00,even odd rule,line width=0.480pt]
        (660.6152,-1654.8398) ellipse (2.0320cm and 0.5080cm);
    \path[draw=black,line join=miter,line cap=butt,miter limit=4.00,even odd
      rule,line width=0.640pt] (-1600.0001,1292.3621) circle (0.9031cm);
    \path[cm={{0.64439,-0.7647,0.64439,0.7647,(0.0,0.0)}},draw=black,line
      join=miter,line cap=butt,miter limit=4.00,even odd rule,line width=0.384pt]
      (-1940.8695,-293.8089) ellipse (1.6255cm and 0.4064cm);
    \path[cm={{0.99996,0.00934,-0.09953,0.99503,(0.0,0.0)}},draw=black,line
      join=miter,line cap=butt,miter limit=4.00,even odd rule,line width=0.450pt]
      (-1501.3894,1450.7275) ellipse (1.8025cm and 0.5040cm);
    \path[cm={{-0.45078,-0.89264,0.78843,-0.61513,(0.0,0.0)}},draw=black,line
      join=miter,line cap=butt,miter limit=4.00,even odd rule,line width=0.505pt]
      (-132.7269,-2268.3213) ellipse (1.6103cm and 0.7093cm);
    \path[cm={{-0.62301,-0.78221,0.8731,-0.48755,(0.0,0.0)}},draw=black,line
      join=miter,line cap=butt,miter limit=4.00,even odd rule,line width=0.281pt]
      (-684.8889,-2571.6060) ellipse (1.1489cm and 0.3073cm);
    \path[draw=black,line join=miter,line cap=butt,miter limit=4.00,even odd
      rule,line width=0.640pt] (-1551.4287,1423.7908) circle (0.9031cm);
    \path[draw=black,line join=miter,line cap=butt,miter limit=4.00,even odd
      rule,line width=0.640pt] (-1740.0001,1423.7906) circle (0.9031cm);
    \path[draw=black,line join=miter,line cap=butt,miter limit=4.00,even odd
      rule,line width=0.640pt] (-1685.7145,1578.0764) circle (0.9031cm);
    \path[cm={{0.77315,-0.63423,0.77315,0.63423,(0.0,0.0)}},draw=black,line
      join=miter,line cap=butt,miter limit=4.00,even odd rule,line width=0.665pt]
      (-2458.3303,135.7292) ellipse (2.8155cm and 0.7039cm);
    \path[cm={{-0.97834,-0.207,0.27782,-0.96063,(0.0,0.0)}},draw=black,line
      join=miter,line cap=butt,miter limit=4.00,even odd rule,line width=0.414pt]
      (1119.1946,-2193.2358) ellipse (1.5210cm and 0.5039cm);
    \path[cm={{0.40788,0.91304,-0.85255,0.52264,(0.0,0.0)}},draw=black,line
      join=miter,line cap=butt,miter limit=4.00,even odd rule,line width=0.449pt]
      (719.8888,2420.5334) ellipse (1.4348cm and 0.6297cm);
    \path[scale=-1.000,draw=black,line join=miter,line cap=butt,miter
      limit=4.00,even odd rule,line width=0.640pt] (1625.7144,-1886.6479) circle
      (0.9031cm);
    \path[scale=-1.000,draw=black,line join=miter,line cap=butt,miter
      limit=4.00,even odd rule,line width=0.640pt] (1774.2858,-1843.7908) circle
      (0.9031cm);
    \path[draw=black,line join=miter,line cap=butt,miter limit=4.00,even odd
      rule,line width=0.640pt] (-1700.0002,1726.6478) circle (0.9031cm);
    \path[cm={{0.36273,-0.93189,0.96869,0.24829,(0.0,0.0)}},draw=black,line
      join=miter,line cap=butt,miter limit=4.00,even odd rule,line width=0.430pt]
      (-2356.4368,-722.1372) ellipse (1.5236cm and 0.5442cm);
    \path[cm={{-0.7825,-0.62265,0.87864,-0.47748,(0.0,0.0)}},draw=black,line
      join=miter,line cap=butt,miter limit=4.00,even odd rule,line width=0.257pt]
      (-449.5538,-2156.3132) ellipse (0.9230cm and 0.3201cm);
    \path[draw=black,line join=miter,line cap=butt,miter limit=4.00,even odd
      rule,line width=0.640pt] (-1514.2859,1949.5051) circle (0.9031cm);
    \path[draw=black,line join=miter,line cap=butt,miter limit=4.00,even odd
      rule,line width=0.640pt] (-1477.1429,1338.0764) circle (0.9031cm);
    \path[draw=black,line join=miter,line cap=butt,miter limit=4.00,even odd
      rule,line width=0.640pt] (-1328.5714,1989.5051) circle (0.9031cm);
    \path[cm={{0.89985,0.43621,-0.43621,0.89985,(0.0,0.0)}},draw=black,line
      join=miter,line cap=butt,miter limit=4.00,even odd rule,line width=0.480pt]
      (-411.1078,2400.7046) ellipse (2.0320cm and 0.5080cm);
    \path[scale=-1.000,draw=black,line join=miter,line cap=butt,miter
      limit=4.00,even odd rule,line width=0.640pt] (1022.8572,-1706.6478) circle
      (0.9031cm);
    \path[scale=-1.000,draw=black,line join=miter,line cap=butt,miter
      limit=4.00,even odd rule,line width=0.640pt] (1207.7362,-2038.0767) circle
      (0.9031cm);
    \path[cm={{-0.82498,0.56516,-0.82498,-0.56516,(0.0,0.0)}},draw=black,line
      join=miter,line cap=butt,miter limit=4.00,even odd rule,line width=0.178pt]
      (2131.3291,-301.9961) ellipse (0.7527cm and 0.1882cm);
    \path[cm={{-0.99899,-0.04483,-0.14229,-0.98982,(0.0,0.0)}},draw=black,line
      join=miter,line cap=butt,miter limit=4.00,even odd rule,line width=0.417pt]
      (1346.3007,-1938.1511) ellipse (1.4854cm and 0.5238cm);
    \path[cm={{-0.10234,0.99475,-0.85092,0.5253,(0.0,0.0)}},draw=black,line
      join=miter,line cap=butt,miter limit=4.00,even odd rule,line width=0.497pt]
      (1306.6970,1228.5986) ellipse (1.8865cm and 0.5856cm);
    \path[cm={{0.61676,0.78715,-0.86966,0.49365,(0.0,0.0)}},draw=black,line
      join=miter,line cap=butt,miter limit=4.00,even odd rule,line width=0.327pt]
      (903.5187,1871.8120) ellipse (1.3428cm and 0.3570cm);
    \path[scale=-1.000,draw=black,line join=miter,line cap=butt,miter
      limit=4.00,even odd rule,line width=0.640pt] (1162.0220,-1855.2192) circle
      (0.9031cm);
    \path[cm={{-0.70711,0.70711,-0.70711,-0.70711,(0.0,0.0)}},draw=black,line
      join=miter,line cap=butt,miter limit=4.00,even odd rule,line width=0.480pt]
      (2030.4762,-488.1443) ellipse (2.0320cm and 0.5080cm);
    \path[cm={{0.81106,-0.58497,0.07316,0.99732,(0.0,0.0)}},draw=black,line
      join=miter,line cap=butt,miter limit=4.00,even odd rule,line width=0.205pt]
      (-2205.1895,167.1333) ellipse (0.7222cm and 0.2618cm);
    \path[cm={{-0.62514,-0.78052,0.21132,-0.97742,(0.0,0.0)}},draw=black,line
      join=miter,line cap=butt,miter limit=4.00,even odd rule,line width=0.341pt]
      (1029.5027,-2374.9661) ellipse (0.8986cm and 0.5790cm);
    \path[draw=black,line join=miter,line cap=butt,miter limit=4.00,even odd
      rule,line width=0.640pt] (-1193.4504,1466.6477) circle (0.9031cm);
    \path[draw=black,line join=miter,line cap=butt,miter limit=4.00,even odd
      rule,line width=0.640pt] (-1076.3077,1458.0762) circle (0.9031cm);
    \path[draw=black,line join=miter,line cap=butt,miter limit=4.00,even odd
      rule,line width=0.640pt] (-1116.3076,1578.0763) circle (0.9031cm);
    \path[scale=-1.000,draw=black,line join=miter,line cap=butt,miter
      limit=4.00,even odd rule,line width=0.640pt] (1207.7362,-1298.0763) circle
      (0.9031cm);
    \path[cm={{0.23861,0.97112,-0.97112,0.23861,(0.0,0.0)}},draw=black,line
      join=miter,line cap=butt,miter limit=4.00,even odd rule,line width=0.480pt]
      (1065.2393,1475.9700) ellipse (2.0320cm and 0.5080cm);
    \path[cm={{-0.69738,0.7167,-0.54532,-0.83823,(0.0,0.0)}},draw=black,line
      join=miter,line cap=butt,miter limit=4.00,even odd rule,line width=0.284pt]
      (1832.1473,-36.6036) ellipse (1.0088cm and 0.3581cm);
    \path[cm={{0.7825,0.62265,-0.87864,0.47748,(0.0,0.0)}},draw=black,line
      join=miter,line cap=butt,miter limit=4.00,even odd rule,line width=0.257pt]
      (1272.6327,2572.9548) ellipse (0.9230cm and 0.3201cm);
    \path[scale=-1.000,draw=black,line join=miter,line cap=butt,miter
      limit=4.00,even odd rule,line width=0.640pt] (1293.4504,-1400.0763) circle
      (0.9031cm);
    \path[cm={{-0.89985,-0.43621,0.43621,-0.89985,(0.0,0.0)}},draw=black,line
      join=miter,line cap=butt,miter limit=4.00,even odd rule,line width=0.480pt]
      (662.6527,-1820.9352) ellipse (2.0320cm and 0.5080cm);
  \end{scope}
  \path[fill=black,line join=miter,line cap=butt,line width=0.800pt]
    (320.4565,2514.3599) node[above right] (text4474-6-6) {(d)};
  \path[draw=black,line join=miter,line cap=butt,miter limit=4.00,even odd
    rule,line width=0.640pt] (257.1427,1152.3621) circle (0.9031cm);
  \path[cm={{0.70711,-0.70711,0.70711,0.70711,(0.0,0.0)}},draw=black,line
    join=miter,line cap=butt,miter limit=4.00,even odd rule,line width=0.480pt]
    (-736.0512,994.6502) ellipse (2.0320cm and 0.5080cm);
  \path[cm={{0.97557,-0.21967,0.14792,0.989,(0.0,0.0)}},draw=black,line
    join=miter,line cap=butt,miter limit=4.00,even odd rule,line width=0.414pt]
    (-170.5541,1289.0756) ellipse (1.5210cm and 0.5039cm);
  \path[cm={{-0.05183,-0.99866,0.98513,-0.17183,(0.0,0.0)}},draw=black,line
    join=miter,line cap=butt,miter limit=4.00,even odd rule,line width=0.356pt]
    (-1384.5962,-130.8490) ellipse (1.1146cm and 0.5086cm);
  \path[cm={{-0.61729,-0.78673,0.86996,-0.49312,(0.0,0.0)}},draw=black,line
    join=miter,line cap=butt,miter limit=4.00,even odd rule,line width=0.401pt]
    (-1406.5737,-1170.4763) ellipse (1.6452cm and 0.4376cm);
  \path[draw=black,line join=miter,line cap=butt,miter limit=4.00,even odd
    rule,line width=0.640pt] (108.5713,1292.3622) circle (0.9031cm);
  \path[draw=black,line join=miter,line cap=butt,miter limit=4.00,even odd
    rule,line width=0.640pt] (-60.0001,1332.3621) circle (0.9031cm);
  \path[draw=black,line join=miter,line cap=butt,miter limit=4.00,even odd
    rule,line width=0.640pt] (-200.0001,1620.9335) circle (0.9031cm);
  \path[draw=black,line join=miter,line cap=butt,miter limit=4.00,even odd
    rule,line width=0.640pt] (-54.2858,1475.2192) circle (0.9031cm);
  \path[cm={{0.70711,-0.70711,0.70711,0.70711,(0.0,0.0)}},draw=black,line
    join=miter,line cap=butt,miter limit=4.00,even odd rule,line width=0.480pt]
    (-1180.5183,1006.7721) ellipse (2.0320cm and 0.5080cm);
  \path[cm={{-0.66187,-0.74962,0.79591,-0.60542,(0.0,0.0)}},draw=black,line
    join=miter,line cap=butt,miter limit=4.00,even odd rule,line width=0.414pt]
    (-1534.0876,-1320.6155) ellipse (1.5210cm and 0.5039cm);
  \path[cm={{0.05183,0.99866,-0.98513,0.17183,(0.0,0.0)}},draw=black,line
    join=miter,line cap=butt,miter limit=4.00,even odd rule,line width=0.356pt]
    (1789.2892,189.8433) ellipse (1.1146cm and 0.5086cm);
  \path[scale=-1.000,draw=black,line join=miter,line cap=butt,miter
    limit=4.00,even odd rule,line width=0.640pt] (-39.9999,-2009.5050) circle
    (0.9031cm);
  \path[scale=-1.000,draw=black,line join=miter,line cap=butt,miter
    limit=4.00,even odd rule,line width=0.640pt] (91.4285,-1892.3623) circle
    (0.9031cm);
  \path[scale=-1.000,draw=black,line join=miter,line cap=butt,miter
    limit=4.00,even odd rule,line width=0.640pt] (97.1429,-1749.5050) circle
    (0.9031cm);
  \path[draw=black,line join=miter,line cap=butt,miter limit=4.00,even odd
    rule,line width=0.640pt] (102.8570,2200.9336) circle (0.9031cm);
  \path[cm={{-0.23861,-0.97112,0.97112,-0.23861,(0.0,0.0)}},draw=black,line
    join=miter,line cap=butt,miter limit=4.00,even odd rule,line width=0.480pt]
    (-2060.0684,-432.6140) ellipse (2.0320cm and 0.5080cm);
  \path[cm={{0.63996,-0.7684,0.4865,0.87368,(0.0,0.0)}},draw=black,line
    join=miter,line cap=butt,miter limit=4.00,even odd rule,line width=0.228pt]
    (-987.8532,1593.1045) ellipse (0.7970cm and 0.2911cm);
  \path[cm={{-0.7825,-0.62265,0.87864,-0.47748,(0.0,0.0)}},draw=black,line
    join=miter,line cap=butt,miter limit=4.00,even odd rule,line width=0.257pt]
    (-1279.0530,-781.3988) ellipse (0.9230cm and 0.3201cm);
  \path[draw=black,line join=miter,line cap=butt,miter limit=4.00,even odd
    rule,line width=0.640pt] (185.7142,2109.5051) circle (0.9031cm);
  \path[draw=black,line join=miter,line cap=butt,miter limit=4.00,even odd
    rule,line width=0.640pt] (379.9999,1198.0764) circle (0.9031cm);
  \path[draw=black,line join=miter,line cap=butt,miter limit=4.00,even odd
    rule,line width=0.640pt] (371.4286,2149.5051) circle (0.9031cm);
  \path[cm={{0.89985,0.43621,-0.43621,0.89985,(0.0,0.0)}},draw=black,line
    join=miter,line cap=butt,miter limit=4.00,even odd rule,line width=0.480pt]
    (1188.4227,1803.1241) ellipse (2.0320cm and 0.5080cm);
  \path[scale=-1.000,draw=black,line join=miter,line cap=butt,miter
    limit=4.00,even odd rule,line width=0.640pt] (-939.9999,-1729.5050) circle
    (0.9031cm);
  \path[scale=-1.000,draw=black,line join=miter,line cap=butt,miter
    limit=4.00,even odd rule,line width=0.640pt] (-492.2638,-2198.0767) circle
    (0.9031cm);
  \path[cm={{-0.70711,0.70711,-0.70711,-0.70711,(0.0,0.0)}},draw=black,line
    join=miter,line cap=butt,miter limit=4.00,even odd rule,line width=0.480pt]
    (1103.1562,-1904.3781) ellipse (2.0320cm and 0.5080cm);
  \path[cm={{-0.97557,0.21967,-0.14792,-0.989,(0.0,0.0)}},draw=black,line
    join=miter,line cap=butt,miter limit=4.00,even odd rule,line width=0.414pt]
    (-416.7930,-2153.3213) ellipse (1.5210cm and 0.5039cm);
  \path[cm={{0.05183,0.99866,-0.98513,0.17183,(0.0,0.0)}},draw=black,line
    join=miter,line cap=butt,miter limit=4.00,even odd rule,line width=0.356pt]
    (2069.9702,-709.8260) ellipse (1.1146cm and 0.5086cm);
  \path[cm={{0.61676,0.78715,-0.86966,0.49365,(0.0,0.0)}},draw=black,line
    join=miter,line cap=butt,miter limit=4.00,even odd rule,line width=0.327pt]
    (1903.3331,323.8407) ellipse (1.3428cm and 0.3570cm);
  \path[scale=-1.000,draw=black,line join=miter,line cap=butt,miter
    limit=4.00,even odd rule,line width=0.640pt] (-640.8354,-2068.0764) circle
    (0.9031cm);
  \path[scale=-1.000,draw=black,line join=miter,line cap=butt,miter
    limit=4.00,even odd rule,line width=0.640pt] (-809.4067,-2018.0764) circle
    (0.9031cm);
  \path[scale=-1.000,draw=black,line join=miter,line cap=butt,miter
    limit=4.00,even odd rule,line width=0.640pt] (-803.6923,-1875.2192) circle
    (0.9031cm);
  \path[cm={{-0.70711,0.70711,-0.70711,-0.70711,(0.0,0.0)}},draw=black,line
    join=miter,line cap=butt,miter limit=4.00,even odd rule,line width=0.480pt]
    (658.6890,-1892.2562) ellipse (2.0320cm and 0.5080cm);
  \path[cm={{0.66187,0.74962,-0.79591,0.60542,(0.0,0.0)}},draw=black,line
    join=miter,line cap=butt,miter limit=4.00,even odd rule,line width=0.414pt]
    (1594.5883,339.6053) ellipse (1.5210cm and 0.5039cm);
  \path[cm={{-0.05183,-0.99866,0.98513,-0.17183,(0.0,0.0)}},draw=black,line
    join=miter,line cap=butt,miter limit=4.00,even odd rule,line width=0.356pt]
    (-1665.2773,768.8201) ellipse (1.1146cm and 0.5086cm);
  \path[draw=black,line join=miter,line cap=butt,miter limit=4.00,even odd
    rule,line width=0.640pt] (709.4067,1340.9335) circle (0.9031cm);
  \path[draw=black,line join=miter,line cap=butt,miter limit=4.00,even odd
    rule,line width=0.640pt] (840.8351,1458.0762) circle (0.9031cm);
  \path[draw=black,line join=miter,line cap=butt,miter limit=4.00,even odd
    rule,line width=0.640pt] (846.5495,1600.9335) circle (0.9031cm);
  \path[scale=-1.000,draw=black,line join=miter,line cap=butt,miter
    limit=4.00,even odd rule,line width=0.640pt] (-646.5495,-1149.5049) circle
    (0.9031cm);
  \path[cm={{0.23861,0.97112,-0.97112,0.23861,(0.0,0.0)}},draw=black,line
    join=miter,line cap=butt,miter limit=4.00,even odd rule,line width=0.480pt]
    (1372.4104,-360.9366) ellipse (2.0320cm and 0.5080cm);
  \path[cm={{-0.63996,0.7684,-0.4865,-0.87368,(0.0,0.0)}},draw=black,line
    join=miter,line cap=butt,miter limit=4.00,even odd rule,line width=0.228pt]
    (57.4856,-1322.3760) ellipse (0.7970cm and 0.2911cm);
  \path[cm={{0.7825,0.62265,-0.87864,0.47748,(0.0,0.0)}},draw=black,line
    join=miter,line cap=butt,miter limit=4.00,even odd rule,line width=0.257pt]
    (2306.9307,1559.2761) ellipse (0.9230cm and 0.3201cm);
  \path[scale=-1.000,draw=black,line join=miter,line cap=butt,miter
    limit=4.00,even odd rule,line width=0.640pt] (-569.4067,-1245.7905) circle
    (0.9031cm);
  \path[cm={{-0.89985,-0.43621,0.43621,-0.89985,(0.0,0.0)}},draw=black,line
    join=miter,line cap=butt,miter limit=4.00,even odd rule,line width=0.480pt]
    (-947.4193,-884.8541) ellipse (2.0320cm and 0.5080cm);
  \path[draw=black,line join=miter,line cap=butt,miter limit=4.00,even odd
    rule,line width=0.640pt] (182.8570,1300.9335) circle (0.9031cm);
  \path[cm={{0.64439,-0.7647,0.64439,0.7647,(0.0,0.0)}},draw=black,line
    join=miter,line cap=butt,miter limit=4.00,even odd rule,line width=0.384pt]
    (-563.1017,1095.1678) ellipse (1.6255cm and 0.4064cm);
  \path[cm={{0.99996,0.00934,-0.09953,0.99503,(0.0,0.0)}},draw=black,line
    join=miter,line cap=butt,miter limit=4.00,even odd rule,line width=0.450pt]
    (280.7382,1442.6182) ellipse (1.8025cm and 0.5040cm);
  \path[cm={{-0.45078,-0.89264,0.78843,-0.61513,(0.0,0.0)}},draw=black,line
    join=miter,line cap=butt,miter limit=4.00,even odd rule,line width=0.505pt]
    (-1257.4724,-650.1014) ellipse (1.6103cm and 0.7093cm);
  \path[cm={{-0.62301,-0.78221,0.8731,-0.48755,(0.0,0.0)}},draw=black,line
    join=miter,line cap=butt,miter limit=4.00,even odd rule,line width=0.281pt]
    (-1573.4182,-1163.6356) ellipse (1.1489cm and 0.3073cm);
  \path[draw=black,line join=miter,line cap=butt,miter limit=4.00,even odd
    rule,line width=0.640pt] (231.4284,1432.3622) circle (0.9031cm);
  \path[draw=black,line join=miter,line cap=butt,miter limit=4.00,even odd
    rule,line width=0.640pt] (42.8570,1432.3621) circle (0.9031cm);
  \path[draw=black,line join=miter,line cap=butt,miter limit=4.00,even odd
    rule,line width=0.640pt] (97.1426,1586.6478) circle (0.9031cm);
  \path[cm={{0.77315,-0.63423,0.77315,0.63423,(0.0,0.0)}},draw=black,line
    join=miter,line cap=butt,miter limit=4.00,even odd rule,line width=0.665pt]
    (-1312.1019,1295.4724) ellipse (2.8155cm and 0.7039cm);
  \path[cm={{-0.97834,-0.207,0.27782,-0.96063,(0.0,0.0)}},draw=black,line
    join=miter,line cap=butt,miter limit=4.00,even odd rule,line width=0.414pt]
    (-600.4392,-1831.6062) ellipse (1.5210cm and 0.5039cm);
  \path[cm={{0.40788,0.91304,-0.85255,0.52264,(0.0,0.0)}},draw=black,line
    join=miter,line cap=butt,miter limit=4.00,even odd rule,line width=0.449pt]
    (1666.9658,782.4335) ellipse (1.4348cm and 0.6297cm);
  \path[scale=-1.000,draw=black,line join=miter,line cap=butt,miter
    limit=4.00,even odd rule,line width=0.640pt] (-157.1428,-1895.2194) circle
    (0.9031cm);
  \path[scale=-1.000,draw=black,line join=miter,line cap=butt,miter
    limit=4.00,even odd rule,line width=0.640pt] (-8.5713,-1852.3622) circle
    (0.9031cm);
  \path[draw=black,line join=miter,line cap=butt,miter limit=4.00,even odd
    rule,line width=0.640pt] (82.8569,1735.2192) circle (0.9031cm);
  \path[cm={{0.36273,-0.93189,0.96869,0.24829,(0.0,0.0)}},draw=black,line
    join=miter,line cap=butt,miter limit=4.00,even odd rule,line width=0.430pt]
    (-1918.9171,954.5186) ellipse (1.5236cm and 0.5442cm);
  \path[cm={{-0.7825,-0.62265,0.87864,-0.47748,(0.0,0.0)}},draw=black,line
    join=miter,line cap=butt,miter limit=4.00,even odd rule,line width=0.257pt]
    (-1382.3116,-957.9043) ellipse (0.9230cm and 0.3201cm);
  \path[draw=black,line join=miter,line cap=butt,miter limit=4.00,even odd
    rule,line width=0.640pt] (268.5712,1958.0765) circle (0.9031cm);
  \path[draw=black,line join=miter,line cap=butt,miter limit=4.00,even odd
    rule,line width=0.640pt] (305.7141,1346.6478) circle (0.9031cm);
  \path[draw=black,line join=miter,line cap=butt,miter limit=4.00,even odd
    rule,line width=0.640pt] (454.2857,1998.0765) circle (0.9031cm);
  \path[cm={{0.89985,0.43621,-0.43621,0.89985,(0.0,0.0)}},draw=black,line
    join=miter,line cap=butt,miter limit=4.00,even odd rule,line width=0.480pt]
    (1196.9268,1630.7189) ellipse (2.0320cm and 0.5080cm);
  \path[scale=-1.000,draw=black,line join=miter,line cap=butt,miter
    limit=4.00,even odd rule,line width=0.640pt] (-759.9999,-1715.2192) circle
    (0.9031cm);
  \path[scale=-1.000,draw=black,line join=miter,line cap=butt,miter
    limit=4.00,even odd rule,line width=0.640pt] (-575.1209,-2046.6481) circle
    (0.9031cm);
  \path[cm={{-0.82498,0.56516,-0.82498,-0.56516,(0.0,0.0)}},draw=black,line
    join=miter,line cap=butt,miter limit=4.00,even odd rule,line width=0.178pt]
    (1058.3679,-1390.1237) ellipse (0.7527cm and 0.1882cm);
  \path[cm={{-0.99899,-0.04483,-0.14229,-0.98982,(0.0,0.0)}},draw=black,line
    join=miter,line cap=butt,miter limit=4.00,even odd rule,line width=0.417pt]
    (-448.6964,-1865.5155) ellipse (1.4854cm and 0.5238cm);
  \path[cm={{-0.10234,0.99475,-0.85092,0.5253,(0.0,0.0)}},draw=black,line
    join=miter,line cap=butt,miter limit=4.00,even odd rule,line width=0.497pt]
    (2497.3652,-1009.8248) ellipse (1.8865cm and 0.5856cm);
  \path[cm={{0.61676,0.78715,-0.86966,0.49365,(0.0,0.0)}},draw=black,line
    join=miter,line cap=butt,miter limit=4.00,even odd rule,line width=0.327pt]
    (1800.9285,458.1929) ellipse (1.3428cm and 0.3570cm);
  \path[scale=-1.000,draw=black,line join=miter,line cap=butt,miter
    limit=4.00,even odd rule,line width=0.640pt] (-620.8351,-1863.7906) circle
    (0.9031cm);
  \path[cm={{-0.70711,0.70711,-0.70711,-0.70711,(0.0,0.0)}},draw=black,line
    join=miter,line cap=butt,miter limit=4.00,even odd rule,line width=0.480pt]
    (775.8668,-1754.8755) ellipse (2.0320cm and 0.5080cm);
  \path[cm={{0.81106,-0.58497,0.07316,0.99732,(0.0,0.0)}},draw=black,line
    join=miter,line cap=butt,miter limit=4.00,even odd rule,line width=0.205pt]
    (-118.1949,1399.8302) ellipse (0.7222cm and 0.2618cm);
  \path[cm={{-0.62514,-0.78052,0.21132,-0.97742,(0.0,0.0)}},draw=black,line
    join=miter,line cap=butt,miter limit=4.00,even odd rule,line width=0.341pt]
    (-1218.5729,-588.5378) ellipse (0.8986cm and 0.5790cm);
  \path[draw=black,line join=miter,line cap=butt,miter limit=4.00,even odd
    rule,line width=0.640pt] (589.4067,1475.2191) circle (0.9031cm);
  \path[draw=black,line join=miter,line cap=butt,miter limit=4.00,even odd
    rule,line width=0.640pt] (706.5494,1466.6476) circle (0.9031cm);
  \path[draw=black,line join=miter,line cap=butt,miter limit=4.00,even odd
    rule,line width=0.640pt] (666.5495,1586.6477) circle (0.9031cm);
  \path[scale=-1.000,draw=black,line join=miter,line cap=butt,miter
    limit=4.00,even odd rule,line width=0.640pt] (-575.1209,-1306.6477) circle
    (0.9031cm);
  \path[cm={{0.23861,0.97112,-0.97112,0.23861,(0.0,0.0)}},draw=black,line
    join=miter,line cap=butt,miter limit=4.00,even odd rule,line width=0.480pt]
    (1498.9655,-253.3462) ellipse (2.0320cm and 0.5080cm);
  \path[cm={{-0.69738,0.7167,-0.54532,-0.83823,(0.0,0.0)}},draw=black,line
    join=miter,line cap=butt,miter limit=4.00,even odd rule,line width=0.284pt]
    (304.8047,-1352.7469) ellipse (1.0088cm and 0.3581cm);
  \path[cm={{0.7825,0.62265,-0.87864,0.47748,(0.0,0.0)}},draw=black,line
    join=miter,line cap=butt,miter limit=4.00,even odd rule,line width=0.257pt]
    (2205.3906,1374.5459) ellipse (0.9230cm and 0.3201cm);
  \path[scale=-1.000,draw=black,line join=miter,line cap=butt,miter
    limit=4.00,even odd rule,line width=0.640pt] (-489.4067,-1408.6477) circle
    (0.9031cm);
  \path[cm={{-0.89985,-0.43621,0.43621,-0.89985,(0.0,0.0)}},draw=black,line
    join=miter,line cap=butt,miter limit=4.00,even odd rule,line width=0.480pt]
    (-945.3818,-1050.9495) ellipse (2.0320cm and 0.5080cm);
  \path[cm={{0.6985,-0.71561,0.66274,0.74885,(0.0,0.0)}},draw=black,line
    join=miter,line cap=butt,miter limit=4.00,even odd rule,line width=0.414pt]
    (-1008.1724,1288.9022) ellipse (1.5210cm and 0.5039cm);
  \path[cm={{-0.02942,-0.99957,0.9558,-0.29401,(0.0,0.0)}},draw=black,line
    join=miter,line cap=butt,miter limit=4.00,even odd rule,line width=0.477pt]
    (-1580.7015,181.5261) ellipse (1.9560cm and 0.5220cm);
  \path[cm={{-0.61729,-0.78673,0.86996,-0.49312,(0.0,0.0)}},draw=black,line
    join=miter,line cap=butt,miter limit=4.00,even odd rule,line width=0.401pt]
    (-1621.8063,-850.2687) ellipse (1.6452cm and 0.4376cm);
  \path[draw=black,line join=miter,line cap=butt,miter limit=4.00,even odd
    rule,line width=0.640pt] (211.4286,1632.3621) circle (0.9031cm);
  \path[cm={{0.70711,-0.70711,0.70711,0.70711,(0.0,0.0)}},draw=black,line
    join=miter,line cap=butt,miter limit=4.00,even odd rule,line width=0.480pt]
    (-897.6757,1305.7772) ellipse (2.0320cm and 0.5080cm);
  \path[cm={{-0.66187,-0.74962,0.79591,-0.60542,(0.0,0.0)}},draw=black,line
    join=miter,line cap=butt,miter limit=4.00,even odd rule,line width=0.414pt]
    (-1517.7549,-1109.5942) ellipse (1.5210cm and 0.5039cm);
  \path[cm={{-0.36268,0.93191,-0.99339,-0.11482,(0.0,0.0)}},draw=black,line
    join=miter,line cap=butt,miter limit=4.00,even odd rule,line width=0.467pt]
    (1873.7484,-974.5889) ellipse (1.8758cm and 0.5204cm);
  \path[scale=-1.000,draw=black,line join=miter,line cap=butt,miter
    limit=4.00,even odd rule,line width=0.640pt] (-605.7142,-1700.9336) circle
    (0.9031cm);
  \path[scale=-1.000,draw=black,line join=miter,line cap=butt,miter
    limit=4.00,even odd rule,line width=0.640pt] (-314.2857,-1760.9336) circle
    (0.9031cm);
  \path[draw=black,line join=miter,line cap=butt,miter limit=4.00,even odd
    rule,line width=0.640pt] (368.5712,1509.5050) circle (0.9031cm);
  \path[cm={{-0.28187,-0.95945,0.97953,-0.20129,(0.0,0.0)}},draw=black,line
    join=miter,line cap=butt,miter limit=4.00,even odd rule,line width=0.440pt]
    (-1469.8029,-81.6766) ellipse (1.7219cm and 0.5042cm);
  \path[cm={{-0.85486,-0.51886,0.34346,-0.93917,(0.0,0.0)}},draw=black,line
    join=miter,line cap=butt,miter limit=4.00,even odd rule,line width=0.505pt]
    (-964.6484,-1103.2455) ellipse (1.6103cm and 0.7093cm);
  \path[cm={{-0.62301,-0.78221,0.8731,-0.48755,(0.0,0.0)}},draw=black,line
    join=miter,line cap=butt,miter limit=4.00,even odd rule,line width=0.281pt]
    (-1786.8259,-844.6866) ellipse (1.1489cm and 0.3073cm);
  \path[draw=black,line join=miter,line cap=butt,miter limit=4.00,even odd
    rule,line width=0.640pt] (508.5711,1598.0764) circle (0.9031cm);
  \path[cm={{0.71569,-0.69842,0.94845,-0.31694,(0.0,0.0)}},draw=black,line
    join=miter,line cap=butt,miter limit=4.00,even odd rule,line width=0.323pt]
    (-3890.0752,3417.4167) ellipse (1.2456cm and 0.3752cm);
  \path[cm={{-0.99861,0.05272,0.4415,-0.89726,(0.0,0.0)}},draw=black,line
    join=miter,line cap=butt,miter limit=4.00,even odd rule,line width=0.478pt]
    (-1478.9811,-2157.7256) ellipse (1.7881cm and 0.5715cm);
  \path[cm={{0.9998,-0.02003,0.17767,0.98409,(0.0,0.0)}},draw=black,line
    join=miter,line cap=butt,miter limit=4.00,even odd rule,line width=0.486pt]
    (-321.6482,1769.7924) ellipse (1.6754cm and 0.6317cm);
  \path[scale=-1.000,draw=black,line join=miter,line cap=butt,miter
    limit=4.00,even odd rule,line width=0.640pt] (-419.9998,-1863.7908) circle
    (0.9031cm);
  \path[draw=black,line join=miter,line cap=butt,miter limit=4.00,even odd
    rule,line width=0.640pt] (494.2854,1746.6478) circle (0.9031cm);
  \path[scale=-1.000,draw=black,line join=miter,line cap=butt,miter
    limit=4.00,even odd rule,line width=0.640pt] (-405.7143,-1689.5050) circle
    (0.9031cm);
  \path[cm={{0.5989,0.80082,-0.85949,0.51115,(0.0,0.0)}},draw=black,line
    join=miter,line cap=butt,miter limit=4.00,even odd rule,line width=0.319pt]
    (1705.8148,537.7660) ellipse (1.3200cm and 0.3448cm);
  \path[cm={{-0.63701,0.77085,-0.63701,-0.77085,(0.0,0.0)}},draw=black,line
    join=miter,line cap=butt,miter limit=4.00,even odd rule,line width=0.345pt]
    (811.6473,-1528.3491) ellipse (1.4612cm and 0.3653cm);
  \path[cm={{-0.04019,-0.99919,0.9756,-0.21956,(0.0,0.0)}},draw=black,line
    join=miter,line cap=butt,miter limit=4.00,even odd rule,line width=0.405pt]
    (-1900.9943,546.3368) ellipse (1.4350cm and 0.5127cm);
  \path[cm={{0.9807,-0.19552,0.04843,0.99883,(0.0,0.0)}},draw=black,line
    join=miter,line cap=butt,miter limit=4.00,even odd rule,line width=0.327pt]
    (601.0002,1829.1567) ellipse (1.3428cm and 0.3570cm);
  \path[cm={{-0.70711,0.70711,-0.70711,-0.70711,(0.0,0.0)}},draw=black,line
    join=miter,line cap=butt,miter limit=4.00,even odd rule,line width=0.480pt]
    (1121.3389,-1449.8094) ellipse (2.0320cm and 0.5080cm);
  \path[cm={{-0.03438,-0.99941,0.96709,-0.25445,(0.0,0.0)}},draw=black,line
    join=miter,line cap=butt,miter limit=4.00,even odd rule,line width=0.440pt]
    (-1621.1812,463.1957) ellipse (1.6754cm and 0.5166cm);
  \path[cm={{-0.99722,-0.07447,0.04778,-0.99886,(0.0,0.0)}},draw=black,line
    join=miter,line cap=butt,miter limit=4.00,even odd rule,line width=0.430pt]
    (-928.0042,-1650.8502) ellipse (1.6307cm and 0.5085cm);
  \path[cm={{0.89599,0.44408,-0.97656,0.21526,(0.0,0.0)}},draw=black,line
    join=miter,line cap=butt,miter limit=4.00,even odd rule,line width=0.559pt]
    (2688.7085,2155.9055) ellipse (2.0170cm and 0.6937cm);
  \path[cm={{0.68114,0.73215,-0.80612,0.59176,(0.0,0.0)}},draw=black,line
    join=miter,line cap=butt,miter limit=4.00,even odd rule,line width=0.213pt]
    (1659.5432,850.0729) ellipse (0.7858cm and 0.2586cm);

\end{tikzpicture}

\caption{(a) A shell  is collapsing in empty space; in its  classical evolution it would create a horizon when it reached the dotted circle.  (b)  In the theory with fuzzballs, there is a nucleation of `bubbles' as the shell comes close to this dotted circle. Since the shell loses some  energy  in creating these bubbles, the location where the classical horizon would form moves to a smaller radius. (c) The shell keeps moving inwards, losing more and more energy to nucleated bubbles, and thus always staying outside its horizon. (d) We finally get a fuzzball with no horizon or singularity.} \label{fig2}

\end{figure}
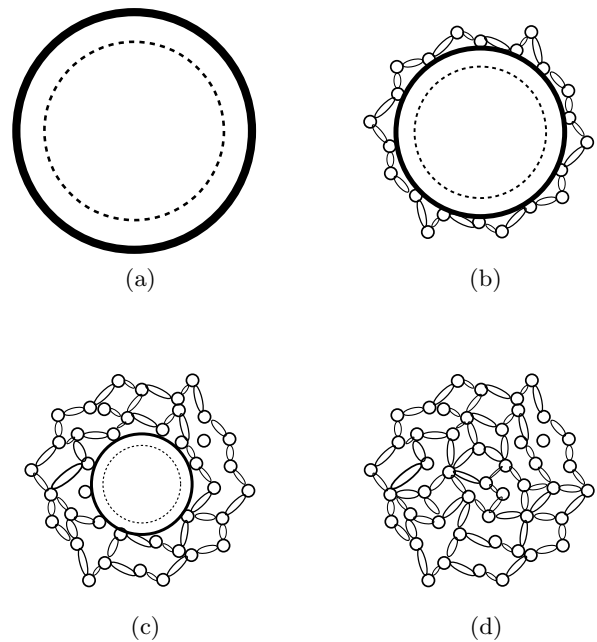

Now consider a spherical shell of mass $M$, collapsing radially inwards, with no other matter present. We have the following picture (we draw the steps schematically in fig.\ref{fig2}):

\mm

(a) When the shell is far from its horizon radius $r_h=2GM$, the motion of the shell is given by semiclassical physics (fig.\ref{fig2}(a)). 

\mm

(b)  When the shell reaches $r=r_h+\epsilon$, with $\epsilon\ll r_h$, there is a nucleation of  `bubbles' just outside the location of the shell. (There can always be quantum fluctuations creating such bubbles, but they become `less expensive' near the shell because of the large redshift when the shell is near its horizon radius.)
The bubbles cost energy, and this energy is drawn from the shell by the process similar to the process  of backreaction in pair creation. So the shell now has a lower  energy, which we write as  $M-\delta M$. This energy corresponds to a horizon radius   $r=2G(M-\delta M)\equiv r_h-\delta r_h$ (fig.\ref{fig2}(b)). 

\mm

(c) The shell therefore travels a little further inwards without forming a horizon. As it approaches the radius
$r=r_h-\delta r_h$, there is again a nucleation of bubbles. The shell loses some more energy, and so travels further inwards without creating a horizon (fig.\ref{fig2}(c)).

\mm

(d) The shell loses all its energy to the creation of bubbles by the time it reaches $r=0$. The ball shaped region containing all the created bubbles is the `fuzzball': the shell state $|S\rangle$ has transitioned into a linear superposition of fuzzball eigenstates $|F_i\rangle$ (fig.\ref{fig2}(d)). 

\mm

Thus instead of a horizon, we get a horizon sized region filled with a nontrivial structure. This structure is analogous to that found in the fuzzball constructions of \cite{lm4,fuzzballs}, where a ball shaped region is filled with monopoles/antimonopoles, with fluxes/branes wrapped on cycles stretching between the monopole centers. 
Of course the degeneracy of states in our toy model of `Euclidean Schwarzschild bubbles' is not high, but the actual fuzzball states $|F_i\rangle$ of the full string theory are expected to correspond to the $Exp[S_{bek}]$ states of the black hole, and in that case they would indeed describe a vast phase space. The conjecture of \cite{tunnel} is that the large number of states that the shell can tunnel to offsets the small  amplitude for tunneling to any given fuzzball state $|F_i\rangle$, so that we indeed violate the semiclassical approximation and end up with the ball depicted schematically in fig.\ref{fig2}(d). 

\bb

Note that in this picture of fuzzball formation the infalling shell was never trapped inside its own horizon, unlike the shell in the classical picture depicted in fig.\ref{fig1}. Our goal now is to find the physical effects  needed to trigger the nucleation of fuzzballs at $r=2GM+\epsilon$  depicted in fig.\ref{fig2}(b).

\section{\label{sec:6}A picture of what we seek}

In the above section  we have conjectured a picture where a collapsing a shell  of mass $M$ begins to tunnel into fuzzballs just before it reaches the location $r=2GM$. We now come to the central question of this paper: what tells the collapsing shell to change its semiclassical behavior at this location? 

In fig.\ref{fig2} we had depicted the collapse of a shell of mass $M$ in otherwise empty space. For simplicity let us first turn to  the case where we already have a fuzzball of mass $M$, and  a shell of mass $\Delta M$  is coming in from infinity at the speed of light. We will again expect that nucleation of the kind in fig.\ref{fig2}(a) will start when the shell reaches a location 
\be
r\approx 2G(M+\Delta M) +\epsilon
\label{qsemi}
\ee
where $\epsilon$ is small compared to the classical scales in the problem. What tells the incoming shell that its semiclassical motion should be altered at the location (\ref{qsemi})? 

As mentioned in the introduction, we conjecture that the spacetime outside the fuzzball of mass $M$ is not just the traditional vacuum:  there are no particles as such in this region, but the vacuum fluctuations are different from the fluctuations of empty spacetime. In this section we will give toy examples to explain the idea of a region with altered vacuum fluctuations, and simple analogies to explain what such an altered vacuum can do.  

\subsection{\label{sec:6.1}An toy example of an altered vacuum}

Consider an example from electrodynamics, where the Schwinger effect in an electric field replaces the process of pair creation in a gravitational field. Consider two infinite parallel plates, with normal along the $\hat x$ direction. Let one plate (carrying a positive charge density $\sigma$) be located at $x=-L$, and the other plate (carrying a positive charge density $-\sigma$) be located at $x=L$. Between these plates we have an electric field
\be
\vec E = {\sigma\over \epsilon_0}\hat x
\ee
and a potential difference 
\be
V=2L |\vec E|= 2L{\sigma\over \epsilon_0}
\ee
Now suppose the theory contains a scalar field $\phi$ with charge $q$ and mass $m$. A 
particle-antiparticle pair of this scalar field can be produced at a minimum energy cost $\Delta {\cal E} = 2 m c^2$. The positively charged particle moves to the negatively charged plate and the negatively charged particle moves to the positively charged plate; this process generates a drop in energy of $\Delta {\cal E}'= 2q|\vec E| L$. We therefore get a creation of particle pairs if and only if $ \Delta {\cal E}'> \Delta {\cal E}$; i.e., if and only if $L>L_{min}$ where
\be
L_{min} = {mc^2\over q |\vec E|}= {mc^2 \epsilon_0\over q \sigma}
\ee

Let us take $L<L_{min}$, so we do not have any creation of particle pairs. Thus there are no on-shell quanta of the field $\phi$. But this does not mean that the state in region between the plates is the same as it would be in a theory which did {\it not} have the scalar field $\phi$.  To see this, suppose we do an experiment where we place an additional pair of plates, carrying surface charge densities $\pm \t \sigma$ at $x=-\mp \t L$, where $\t L < L$. The field between the plates is now
\be
\vec E = {\sigma+\t \sigma\over \epsilon_0}\hat x
\ee
The potential difference between $x=\pm L$ is now
\be
V  = 2L{\sigma\over \epsilon_0} + 2\t L {\t \sigma\over \epsilon_0}
\ee
We find that the condition $ \Delta {\cal E}'= \Delta {\cal E}$ is now satisfied at
\be
\t L={mc^2\epsilon_0\over q\t \sigma} - L{\sigma\over \t \sigma}\equiv \t L_{ min}
\ee
Thus if we let $\t L>\t L_{min}$ then  we will in fact get creation of particle pairs for the scalar  field $\phi$. 

It may seem that the particle creation we get this way is a small quantum effect, and so nothing dramatic happens when $\t L$ crosses $\t L_{min}$. But now let us add to our toy model a version of `entropy enhancement'.  Instead of one scalar field $\phi$, we take $N$ scalar fields $\phi_i$, with
\be
N\gg 1
\label{qngg}
\ee
We still set $L<L_{min}$. Thus before we add in the extra plates at $x=\pm \t L$, we have no on shell particles. Now we add in  the extra plates at $x=\pm \t L$. When $\t L < \t L_{min}$, there are still no on shell particles of the fields $\phi_i$. But when $\t L$ crosses $\t L_{min}$ we get a large number of created pairs, and the backreaction of this pair creation can create a significant change in the dynamics of the plates at $x=\pm \t L$. 

To summarize, the large number of virtual pairs of the fields $\phi_i$ changes the vacuum between the plates at $x=\pm L$, to a form different from the vacuum in a theory which did not have the fields $\phi_i$.
This is an example of an altered vacuum state, and we have noted that this altered state can lead to a large effect of the dynamics of the plates at $x=\pm \t L$  if we assume  the `entropy enhancement' (\ref{qngg}). This effect is quantum in its origin however, and would be missed if we considered only the classical dynamics of the electrodynamic setup considered  here.

\subsection{\label{sec:6.2}A schematic picture of the near-horizon region}

Let us now ask: what is the consequence of having a region with altered vacuum fluctuations? We illustrate our conjecture with a schematic model:

\mm

(i) Consider the edge of a lake depicted in fig.\ref{fig3}. On the left is land; this represents the interior of the fuzzball $r < r_b$. The water represents the exterior region $r>r_b$. Waves can propagate on the surface of this water, and represent matter quanta in the region $r>r_b$. 

\mm

(ii) If a quantum were travelling in flat spacetime, we would depict it by a wave on a lake with infinite depth. But as we approach $r=r_b$, the vacuum gets altered more and more strongly by the effects depicted in fig.\ref{fig1p}. In our schematic model, the effect of these altered fluctuations is to reduce the depth of the lake; this depth  goes to zero as we reach $r=r_b$. 

\mm

(iii) Consider a shell carrying energy $\Delta M$ which is falling towards the fuzzball surface. This corresponds to a wave on the lake, with the wave height being proportional to $\Delta M$. Since the depth of the lake decreases towards the shore, there will be a point where the height of the wave becomes comparable to the depth of the lake. Beyond this point the will no longer be able to travel freely as if it were a wave on a lake of infinite depth, and we expect new dynamical effects to arise. In the fuzzball, the shell of mass $\Delta M$ will similarly reach some location $r>r_b$ where it is no longer able to proceed as expected by a semiclassical analysis; this is the point where the tunneling into fuzzballs will start to take place. 

\mm

To summarize, we have argued that we should think of  the spacetime outside the fuzzball (i.e. the region $r>r_b$)  as having a `thickness' that reaches zero at the surface of the fuzzball, and increases as we go away from  the fuzzball. Thus it does not  make sense to ask if physics is `normal' outside the fuzzball: the correct question is: the physics is normal for objects upto what energy $\Delta M$? The answer would then be that the physics is normal below an energy where the shell would have started to pass through its own horizon; and at this point we get the entropy-enhanced tunneling which changes the shell into a fuzzball. 

\subsection{\label{sec:6.3}The notion of spacetime having a `thickness'}

The notion that we should associate a `thickness' with spacetime arose in the discussion of \cite{universe}; let us recall this discussion here. 

\begin{figure}
 \includegraphics[scale=.16] {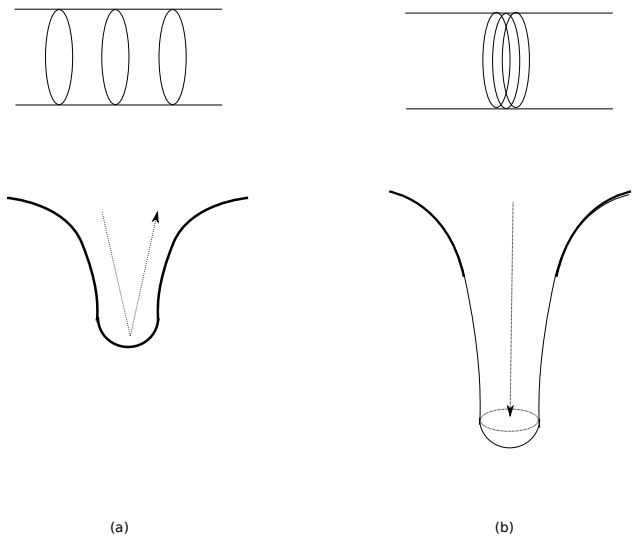}
\caption{\label{fig4} Two microstates of the D1D5 system and their corresponding geometries. (a) The effective string is broken into `singly wound cycles'; the corresponding geometry is a shallow throat. A quantum thrown into this throat  returns back to infinity without creating a horizon. (b) The effective string is `multiply wound'; the corresponding geometry has a deep throat, and a quantum with the same energy will create a fuzzball when it reaches  the location marked by the dotted circle. We say that the extra stretching in (b) has made the space have a smaller `thickness' than the space in (a), so it is depicted with  a thinner line.  }
\end{figure}

Consider the extremal 2-charge hole in string theory given by the D1D5 model.  The D1D5 system is obtained by compactifying IIB string theory as $M_{9,1}\r M_{4,1}\times S^1 \times T^4$. We wrap $n_1$ D1 branes on the $S^1$ and $n_5$ D5 branes on $T^4\times S^1$. The bound state of these branes gives an effective string wound around the $S^1$ with winding number $N=n_1n_5$. This effective string can be partitioned in different ways into `component strings'  with different windings $k$. If the winding of each component string is the same, then the number of component strings is given by $n_c=N/k$. In fig.\ref{fig4} we depict two different partitions: one where all windings are unity, and one where we have a single component string of winding $k\gg 1$.  The corresponding spacetime  solutions  have throats of different depth: the ones with the large winding $k$ has a deeper throat. We say that the geometry has been `stretched' more in this situation. It was then argued that this stretching gives rise to a spacetime which has a lesser `thickness', in the sense that it can be more easily distorted to create a near-extremal black hole. More precisely, it was found in \cite{lm4} that the  black hole threshold is reached when we send into the throat a quantum with enough energy $E$ to excite each component string with its lowest allowed excitation energy. If the $S^1$ has radius $R$, then the lowest excitation  consists of one left and one right moving vibration with energy 
\be
E={1\over k R}+{1\over kR}={2\over kR}
\ee
The minimum total energy required to excite all component strings is then
\be
E_{min}=\left ( {2\over kR}\right ) n_c = {2N\over R} {1\over k^2}
\label{qsimple}
\ee
In the gravity solution, we find that a quantum with energy $E\gtrsim E_{min}$ creates a black hole (fuzzball) instead of just bouncing back from the `cap' at the end of the throat. Note that all parameters like the string coupling $g$ and the size of the $T^4$ cancel out; leaving the simple  expression (\ref{qsimple}) for this critical energy. Since the throats where the spacetime been more stretched  -- the ones with larger $k$ -- are more easily deformed to black holes, we say that the `thickness' of the space decreases when space is stretched. Thus we can think of spacetime as a rubber sheet rather than just a manifold: the thickness of such a rubber sheet decreases when the sheet is stretched, while a manifold has no `thickness' whatsoever.

\subsection{\label{sec:6.4}The $c=1$ Matrix model}

Another system which furnishes  a useful analogy for our purposes is the $c=1$ matrix model. This model does not have a black hole, or the phenomenon of `entropy enhanced tunneling',  so we do not capture all the features of our conjecture. But the model does have a `fermi sea with varying depth', which is similar to fig.\ref{fig3}. 

 The model arises from the quantization of a $N\times N$ Hermitian matrix with a Lagrangian $Tr[\dot M^2 - V(M)]$. The eigenvalues of this matrix behave like fermions, which fill a fermi sea in the ground state \cite{dasjevicki}. Small ripples on this fermi sea are described by a quadratic Lagrangian, and can be mapped to a massless scalar $\phi$. But the depth of the fermi sea goes to zero near its edge.  When a ripple gets close enough to the edge that its height is comparable to the depth of the fermi sea, then a  cubic term becomes relevant in the effective Lagrangian for the scalar, so it no longer behaves as a free field. One can analyze the nonperturbative dynamics that results from this cubic coupling, and regard this as a model for nonperurbative effects in gravity \cite{jevickinonpert}. 
 
 We can map the low energy  behavior of this scalar $\phi$ to a scalar coupled to 1+1 dimensional dilaton gravity.  We then find that the place where the ripple touches the bottom  of the fermi sea is roughly the location where a black hole would have formed if the dilaton gravity description had continued to be valid \cite{pol,jevicki}. 

While the low energy dynamics of the fermi sea resembles dilaton gravity, there is no long-lived black hole type state in the matrix model \cite{kms}. It might therefore seem that this model does not furnish a useful analogy for the information puzzle.  But we should note that  the model arises from an actual quantization of string world sheets, so it is an example of a string theory computation rather than just a schematic model. Thus it is significant that this model leads to a  `fermi sea with varying depth' in a natural way. Further, the breakdown of classical evolution in this model -- resulting from the formation of `folds' on the fermi sea -- can be understood as a nonperturbative effect arising from the interaction between a large number of bosonic quanta \cite{folds}; thus this effect may have some similarities to the idea of `entropy enhancement' that we have used in our picture.\footnote{It should be noted however that the eigenvalues making up the fermi sea decouple from the angular degrees of freedom of the matrix, so a perturbation on the fermi sea is not able to access most of the $\sim N^2$ degrees of freedom in the matrix model.}

\section{\label{sec:7}The `pseudo-Rindler' conjecture}

While the above intuitive examples serve to illustrate the physics we are looking for, they do not  tell us how  a `varying depth sea' should actually arise. String theory is a complete theory which permits no addition of new particles or interactions. So if we wish to argue that the  region  outside a fuzzball is different from the conventional vacuum, then we have to conjecture a concrete source  effects in string theory which can generate the required change of state. This is the question we turn to now.

In this section we explain our conjecture that vacuum fluctuations of the fuzzball change the region outside the fuzzball from Rindler space to what we will call `pseudo-Rindler' space. We will first state the context of problem we are addressing (Section \ref{sec:7.1}), then state our `pseudo-Rindler' conjecture (section \ref{sec:7.2}) and finally use this conjecture to get a picture of infall in a theory with fuzzballs (section \ref{sec:7.3}).

\subsection{\label{sec:7.1}The  state around the horizon}

The traditional understanding of a black hole has been dominated by two ideas:

\mm

(i) The spacetime at the horizon of a large black hole is essentially a part of Minkowski space in its vacuum state. 

\mm

(ii) The region just outside the horizon is locally described by Rindler space; i.e.,   the Schwarzschild coordinates become the Rindler coordinates which cover one quadrant of Mikowski spacetime. 

\mm

It is true that (ii) is implied by (i):    if the region around the horizon is a patch of Minkowski space, then restricting this patch to the part outside the horizon will give Rindler space. But with fuzzballs, we find that (i) is not true: the region $r<2GM+\epsilon$ is altered to a state $|\psi\rangle$ that has very low overlap with the local vacuum (i.e., $\langle 0 |   \psi\rangle \ll 1$). We can now ask if (ii) is still true; i.e., is the space outside the fuzzball locally identical to the Rindler quadrant of empty Minkowski space? Our conjecture will be that with fuzzballs, (ii) is not true either: the vacuum fluctuations in this region change Rindler space to what we will call `pseudo-Rindler' space. 

Before we address how (ii) would be invalidated, let us recall how (i) fails in the fuzzball paradigm. Empty Minkowski space has no mass:
\be
M=0
\label{qtone}
\ee
If we take a black hole whose radius tends to infinity, then we have the limit
\be
M\r \infty
\label{qttwo}
\ee
Classical, it appears that the spacetime generated by the limit (\ref{qttwo}) reproduces, in a patch near the horizon, the locally flat spacetime given by (\ref{qtone}). But in the case (\ref{qtone}) we have a unique ground state; i.e., the number of states is 
\be
{\cal N}=1
\label{qtthree}
\ee
while in the limit (\ref{qttwo}) the number of states goes to infinity
\be
{\cal N} \approx e^{S_{bek}(M)} \r \infty
\label{qtfour}
\ee
In the fuzzball paradigm, the difference between (\ref{qtthree}) and (\ref{qtfour}) prevents us from decoupling a small region around the horizon of a black hole and treating it as a patch of empty Minkowski space. The phase space volume corresponding to  (\ref{qtfour}) grows rapidly and nonlinearly with $M$, and we are forced to look at the complete system as a whole when the tunneling into the fuzzball states $|F_i\rangle$ becomes important. 

\subsection{\label{sec:7.2}Vacuum fluctuations outside the fuzzball}

In fig.\ref{fig1p}(a) we depict the traditional black hole with vacuum at the horizon. Fig.\ref{fig1p}(b) depicts a fuzzball. 

In the region outside the horizon of the traditional hole, we have the same vacuum fluctuations as we would find locally in empty space. Note that the absence of excitations like (\ref{qtfive}) around the black hole prevents us from changing the vacuum; this was the `no-hair theorem' at the quantum level.

But the case is different for the fuzzball, as we see from fig.\ref{fig1p}(b). Now we have a surface at the location $r=2GM+\epsilon$. This surface can emit virtual quanta into the Rindler region, which generates vacuum fluctuations that would be different from those in a patch of empty space. It is these fluctuations which change the spacetime around the hole from Rindler to pseudo-Rindler. The two questions that we must now address are:

\mm

(A) What is the nature of the relevant quantum fluctuations?

\mm

(B) Why should such  fluctuations be important for the dynamics of the hole?

\mm

Let us now state conjecture our answer to these two questions in the fuzzball paradigm. 

\mm

(A') Suppose we start with a fuzzball of mass $M$.  Consider a  fluctuation where the configuration  changes to a fuzzball of mass $M+\Delta M$. Since we do not have the extra energy $\Delta M$, this is a virtual fluctuation, just like the appearance of a virtual  electron-positron pair in the vacuum. Note that the amplitude for this fluctuation will be large if $\Delta M$ is small; i.e., if the hole already has a mass $M$ close to the value $M+\Delta M$. We conjecture that these fluctuations of the hole into fuzzballs of larger size are the fluctuations relevant for changing the polarization of the vacuum in the region outside the fuzzball. 

\mm

(B') Vacuum fluctuations are normally a quantum effect, ignorable for the leading order classical approximation for macroscopic dynamics. But here we encounter the entropy-enhancement effect again: the number of virtual fuzzballs ${\cal N'}$ with mass $M+\Delta M$ is very large
\be
{\cal N'} \approx e^{S_{bek}(M+\Delta M)} \gg 1
\label{ftsix}
\ee
The large number of these virtual fluctuations can compensate the low probability of the fluctuation to any individual fuzzball. Thus the vacuum polarization caused by these fuzzball fluctuations can be significant. The region around the hole polarized by such fluctuations is what we call pseudo-Rindler space, to distinguish it from Rinder space which has just the fluctuations of empty space.

\subsection{\label{sec:7.3}A picture of infall}

Let us use the conjecture above to obtain a picture of infall onto a fuzzball of mass $M$. This fuzzball has a surface at $r_b=2GM+\epsilon$.

\mm

(a)  Start with the fuzzball of mass $M$. Let a shell of mass $\Delta M$ be incident on this fuzzball from infinity.

\mm

(b) When the shell is at large radii $r$ it travels in the usual semiclassical approximation. The fuzzball of mass $M$ has fluctuations to fuzzballs of mass $M+\Delta M$.  But the region near the fuzzball has only energy $M$, so these fluctuations remain virtual.

\mm

(c) When the shell reaches close to $r=2G(M+\Delta M)$, these virtual fluctuations are able to turn into real fluctuations, since now a mass $M+\Delta M$ is available in a region with radius equal to the radius of these virtual fuzzballs. We then get the process outlines in section \ref{sec:3} where the shell breaks up into bubbles, creating a fuzzball state in the region $2GM<r<2G(M+\Delta M)$. At the end of this process we are left with a fuzzball of mass $M+\Delta M$ and radius $2G(M+\Delta M)$. 

\mm

\subsection{\label{sec:7.4}Causality in the collapse process}

Our central question was: how is causality maintained during the process of transitioning to fuzzballs? 
Looking at the transition process conjectured in section \ref{sec:3} we see that we do not have any violation of causality; this is because the shell never gets trapped inside its own horizon. Let us analyze in more detail how such a causality preserving transition is attained in our picture.

Suppose the shell is composed of massless quanta that fall in radially at the speed of light. In this case it is true that the incoming shell cannot  influence the fuzzball surface  at $r=2GM+\epsilon$ when it is still far away from this surface; this is because there has not been time for a light signal to go from the infalling shell to the fuzzball surface. Thus there is certainly no way for the infalling shell at $r>r_b$ to influence the fuzzball surface at $r=2GM+\epsilon$ to send an outwards signal that will change the motion of the shell. So it would seem that the shell would travel inwards while maintaining  in the semiclassical approximation all the way till $r=r_b$ and then crash onto the fuzzball surface. 

But we have argued that this is not the case. The fuzzball (describing the black hole of mass $M$) has been in existence for some time (several crossing times, say).  This allows the mass $M$ to polarize the space outside $r=r_b$ by virtual fluctuations of fuzzballs (fig.\ref{fig1p}(b)),  without any violation of  causality.\footnote{In fact we can make a stronger statement.   The fuzzball was created by matter which fell in from infinity, and this infalling matter crosses all the positions $r>r_b$ in the process of reaching its final location. Thus causality does not forbid this matter from influencing any location $r>r_b$.}   When the shell is at a position $r>r_b$ then it  can react to the altered vacuum polarization at its position $r$.  When the virtual fuzzball fluctuations reaching upto  location $r$ are more massive than the total available energy $M+\Delta M$ then they have very little effect on the infalling shell of mass $\Delta M$. The situation changes at the location $r=2G(M+\Delta M)$, when these virtual fluctuations can turn into on-shell fuzzball states by absorbing the energy $\Delta M$ of the shell.  This change is quite sudden because it involves the competition between two exponentials: a decreasing one from the action required to create the massive fuzzball, and a growing one from the large degeneracy (\ref{ftsix}) of these fuzzballs. In this manner the shell can transition to fuzzballs at a point $r>r_b$ -- before it crosses its own horizon -- and without any violation of causality. 

Given that the spacetime outside the fuzzball has an altered vacuum state, one might wonder if we should say that the fuzzball itself extends past $r=r_b$. The reason that we should not say this is the following. Low energy quanta travel in the region $r>r_b$ just as they would in empty spacetime, so for such quanta the region $r=r_b$ is in fact characterized accurately by the classical metric. In fact for any given mass $\Delta M$ of an infalling shell, the classical metric captures the dynamics to a good approximation for $r>2G(M+\Delta M)$. Thus we should still say that the fuzzball ends at $r=r_b$, but that it alters the vacuum at $r>r_b$.

\section{\label{sec:8c}Comments on the pseudo-Rindler conjecture}

We have proposed that vacuum fluctuations modify the spacetime outside the fuzzball to yield pseudo-Rindler spacetime, and that this effect allows causality to be maintained in the process of fuzzball formation and evaporation. We  now make some observations to explain various aspects of this proposal.

\subsection{\label{sec:8c.1}Collapse in empty space}

In the above discussion, we have started with a fuzzball of mass $M$, and considered a shell of mass $\Delta M$ that collapsed towards the fuzzball.  We can take the limit where $M\r 0$, so that we have just the collapse of a shell in empty space. What does our conjecture about causality say in this case?

The fuzzball of mass $M$ had vacuum fluctuations to fuzzball states of mass $M+\Delta M$. Even if we have no mass in our spacetime, there will be vacuum fluctuations corresponding to fuzzball states of all masses $ M$, centered about all points of spacetime. When a shell of mass $ M$ contracts to the point where it is about to form a horizon, these vacuum fluctuations become converted to `real' fuzzball states, and we get the picture of black hole formation described in fig.\ref{fig2}. This picture preserves causality, as we have noted. 

Note that the tunneling to fuzzball states happens when the matter density in the collapsing shell is still low. As noted in section \ref{sec:1}, this is one of the aspects of the fuzzball paradigm: the large entropy of fuzzball states destroys the semiclassical approximation when the collapsing shell reaches its horizon radius. To understand this a little better let us recall a toy model presented in \cite{kraus}. Consider the collapse of a shell with mass $M$. Imagine that the theory contains a large number $N$ of massless scalar fields $\phi_i$. These fields are in their ground state, so the shell is collapsing in the vacuum. When the shell traverses the region between say $4GM$ to $3GM$, it changes the metric there by order unity, and this deformation creates $\sim 1$ pairs of excitations for each scalar field $\phi_i$. The wavelength of each such created quantum is $\lambda\sim GM$, which corresponds to  an energy $\sim 1/(GM)$. Suppose the number of species $N$ satisfies
\be
N\gg GM^2
\ee
Then the infalling shell will lose its energy to the created quanta before it reaches its horizon, and a black hole will not form. Of course in this toy model $N$ is a fixed number, so we will still get horizon formation if $M\gg (N/G)^\h$. But in a theory with fuzzballs, the number of fuzzball states that we can transition to grows with $M$, and so we do not form a horizon for any $M$.

\subsection{\label{sec:8c.2}Vacuum fluctuations vs. thermal fluctuations}

We have conjectured that the vacuum fluctuations created by the existence of a fuzzball surface at $r\approx 2GM+\epsilon$ polarizes the spacetime outside the fuzzball to a state different from usual empty spacetime. Note that these are {\it quantum} fluctuations rather than thermal fluctuations. To see the difference, consider a an extremal hole with charge $Q=M$. The hole has a temperature $T=0$, so there are no thermal fluctuations near the near horizon. But if we throw in a neutral shell of mass $\Delta M$, then the classical dynamics would create a horizon at  $r_h^{new}-r_h \approx G\sqrt{2Q\Delta M}$. Thus to preserve causality, we would need fuzzball formation at this location $r_h$, which would arise from nontrivial quantum fluctuations at $r^{new}_h$. Thus the fluctuations we are interested in are quantum fluctuations rather than thermal fluctuations. 

It is important to distinguish these vacuum fluctuations caused by the fuzzball from the fluctuations that we get when expressing Minkowski spacetime in Rindler coordinates. Minkowski space of course has its own vacuum fluctuations. If we write Minkowski space is Rindler coordinates, then we are not changing these fluctuations: we are simply splitting them in a different way between what we call the `vacuum' and what we call `particles'. By contrast, when we have a fuzzball boundary at $r\approx 2GM+\epsilon$ then the vacuum fluctuations in the region $r>2GM+\epsilon$ actually change. The presence of the boundary breaks the translation invariance of the local spacetime. This allows new fluctuations of the kind pictured in fig.\ref{fig1p}(b), and it is such fluctuations that help resolve our causality problem. 

Put another way,  we may break up the fluctuations outside the fuzzball into two categories:

\mm

 (i) We have a gas of gravitons near the fuzzball surface; these have the same temperature and energy density as the graviton gas found in  Rindler space.
 
 \mm
 
  (ii) We have excitations that are supported by the fuzzball surface at $r_b=2GM+\epsilon$; these fluctuations are actual deformations of the fuzzball, and would not arise if the fuzzball surface did not break the translation invariance of the local spacetime. 
  
  \mm
  
  It is the fluctuations of  category (ii) which are relevant for creating the entropy enhanced tunneling that we have conjectured to help resolve the causality problem.

\subsection{\label{sec:8c.3}The wavefunctional in superspace}

We have argued that even in empty Minkowski spacetime, there are quantum fluctuations, around every point,  into fuzzball modes of all energies $M$. How should we understand the wavefunctional describing such a vacuum state?

This question was addressed in \cite{kraus}. We should not think of just the Minkowski metric, but of superspace -- the space of all possible metrics. The gravity wavefunctional is a wavefunction on this superspace. Most of the regions of this superspace corresponds to configurations that have a large mass $M>0$; fuzzball solutions are examples of such points in superspace. Empty Minkowski space has $M=0$, so these $M>0$ regions are `under the barrier' for the wavefunction on superspace. The wavefunction in these regions does not vanish; rather it decays towards the direction of larger $M$ just like wavefunctions decay under the barrier in a square-well potential in quantum mechanics.

Ordinarily the part of the wavefunction under the barrier would not be very significant.  But due to the large entropy of fuzzballs the volume of superspace where the wavefunction is under the barrier is very large. Thus most of the wavefunction is actually under the barrier. This part of the wavefunction still does not have a large significance for the propagation of light quanta on spacetime. But now consider the collapse of a shell of mass $M$.  When such an extra energy is available, the part of the wavefunction that was under the barrier at $M=0$ can now become oscillatory. In particular when the shell reaches a radius $r\approx 2GM$, the wavefunction describing fuzzballs of radius $r_b\approx 2GM$ becomes oscillatory. These fuzzballs thus become on-shell states, and this is the tunneling into fuzzballs that we have conjectured. 

It would be interesting to explore further the wavefunctional we have conjectured where a large part is `under the barrier' in the form of fuzzballs. For example, one might ask how the part under the barrier is altered when there is a cosmological constant $\Lambda$, and whether this might favour a small $\Lambda$ over other values. The nature of the vacuum (and in particular its Lorentz invariance) was studied for the Schwinger effect and for bubble nucleation in \cite{vilenkin}.

\subsection{\label{sec:8c.4}Causality in a special limit}

\begin{figure}
 \includegraphics[scale=.22] {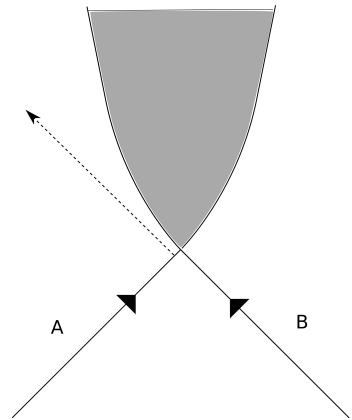}
\caption{\label{figdpcausality} The spacetime generated by the collision of two shock waves \cite{deathpayne}. The shock waves describe massless particles A and B that meet in a head-on collision. The shaded region has a nontrivial structure arising from the effect of the collision. The other regions of the geometry are parts of Minkowski space. The dashed line denotes a light ray emitted by A; such rays can reach infinity for all  emission times upto the collision point. }
\end{figure}

Consider the situation where we make a black hole by a head-on collision of two massless particles. 
One particle A with energy $E$ travels along the positive $x$ axis, while another particle  B with energy $E$ travels along the negative $x$ axis. We can arrange the trajectories so that the particles meet at $x=0, t=0$. 

If signals do not travel faster than the speed of light, then the first particle cannot know if the existence of the second particle until they actually collide at $t=0$. What constraints are set by causality in this situation?

The black hole spacetime  created by  such a collision was studied in \cite{deathpayne}, and is depicted in fig.\ref{figdpcausality}. The shaded region is the interior of the hole; in our paradigm, this is the region where a fuzzball should exist. The other regions are locally flat spacetime. We see that a null ray can start at particle A at any time before the point of collision and escape to infinity; such a ray is depicted by a dashed line in the figure.  A similar situation holds of course for particle B. Thus we see that  causality does not require fuzzball formation before the collision in this example.  After the collision a fuzzball can be generated at the point of impact and expand, maintaining causality, in such a way that its surface lies just outside the shaded region in the figure.

\subsection{\label{sec:8c.5}Fluctuations outside a star}

A black hole of mass $M$ has a horizon radius $r=2GM$, and in the fuzzball paradigm this would be replaced by a fuzzball of radius $r_b=2GM+\epsilon$. We have argued that the region $r>r_b$ contains virtual fluctuations that are important for the dynamics of heavy infalling objects. But now consider a neutron star of mass $M$. The radius of this star is $r_{star}>2GM$. Will here be a similar change in vacuum fluctuations outside this star?

Consider a shell incident on this star with mass $M_{shell}$. 
Then in the classical picture of collapse the shell will form a horizon at a radius 
\be
r_h=2G(M+M_{shell})
\ee
provided that 
\be
r_h>r_{star}
\label{qstar}
\ee
We satisfy the condition (\ref{qstar}) for
\be
M_{shell} > {r_{star}\over 2G} - M
\label{qshell}
\ee
Thus for shells with mass satisfying (\ref{qshell}) 
 we will again have our problem with causality, if we do not have a tunneling to fuzzballs  just before the shell reaches the radius $r_h$. Note that $r_h$ depends on the total mass $M$ in the region interior to the shell, but not on whether this mass $M$ is in the form of a  black hole (fuzzball) or a neutron star. Thus we conjecture that the fluctuations at a radius  $r>r_{star}$ are, to leading order, the  same outside a neutron star and a fuzzball provided the two objects have the same mass. 
 
 But recall that the fluctuations are strong near $r\approx 2GM$ and small further away. Since $r_{star}$ is much larger than $2GM$, the fluctuations at $r>r_{star}$ are quite weak anyway, and cannot be seen unless we probe the system with a shell of mass exceeding (\ref{qshell}). Thus while we do have fluctuations outside a neutron star, small infalling objects  will not be affected in any appreciable way by these fluctuations as they fall onto the surface of the star.

\section{\label{sec:8}Distance and energy scales}

In this section we give estimates for the distance $s$ from the horizon at which an incoming pulse of energy loses semiclassical behavior.  We have conjectured above that this is the location where a horizon would be about to form, so our  estimates are just those that give the location of the horizon for a given energy profile of the infalling object. But it is useful to recall these results in the present context and see the value of $s$ in the Rindler approximation for different kinds of infalling objects. 

\subsection{\label{sec:8.1}The Rindler approximation}

The Schwarzschild hole in $D$ spacetime dimensions is give by the metric
\be
ds^2=-f(r) dt^2 +{dr^2\over f(r)} +r^2d\Omega_{D-2}^2
\label{qfone}
\ee
where
\be
f(r)=1-\left ( {r_0\over r}\right )^{D-3}
\ee
and
\be
r_0= \left ( {16\pi G \over (D-2) \Omega_{D-2}} M \right )^{1\over D-3}
\label{qftwo}
\ee
The near-horizon region is given by $r-r_0\ll r_0$. In this region we define  
\be
s={2\over f'^\h_0}(r-r_0)^\h, ~~~ \tau={f'_0\over 2} t
\ee
where $f'_0=df/dr(r=r_0)$. This gives
\be
ds^2\approx -s^2 d\tau^2 + ds^2 + \sum_{i=1}^{D-2}dx_idx_i
\label{qrindler}
\ee
We see that $s$ measures the distance from the horizon. 

\subsection{\label{sec:8.2}Infalling shell}

Consider the infall of a spherical shell of mass $m\ll M$ onto the hole (\ref{qfone}). From (\ref{qftwo}) we find that the new horizon will form at
\be
s_h\approx {2r_0\over (D-3)} \sqrt{m\over M}
\label{qffive}
\ee
so  the tunneling into fuzzballs will take place just outside this location. It is interesting to note that we cannot take a strict Rindler limit of this process. In such a limit the shell will look like an infinite plane sheet stretching in the directions $x_i$, with some surface energy density $\sigma$. One may then try to ask for the value of $s$ where such a sheet will form a horizon. But as we will now note, the value of $r_0$ does not decouple from such a computation, so we cannot take the limit $r_0\r \infty$. 

Consider the shell as it passes the point $s=s_1$; we assume that  $s_h\ll s_1\ll r_0$. At this point we can set up a local orthonormal frame with unit vectors $\hat t, \hat r$ along the $t,r$ directions. The shell is moving close to the speed of light at this location. The shell will look like a plane with surface energy density in the  local frame given by
\be
\sigma_1 \approx  {m\over \Omega_{D-2} r_0^{D-2}}\left ( s_1  {(D-3)\over 2 r_0}\right ) ^{-1} = {(D-2)\over (D-3)}{1\over 8\pi GM}{m\over s_1} 
\label{qframe}
\ee
where the factor $( s_1  {(D-3)\over 2 r_0} ) ^{-1}$ arises from the redshift at the location $s_1$. One may now try to hold fixed $s_1, \sigma_1$ and ask for the location $s_h$; this would pose the problem in a purely Rindler language. But expressing  (\ref{qffive}) in terms of $\sigma_1$ rather than $m$ we find  that
\be
s_h \approx  r_0 \left ( {32\pi G \sigma_1  s_1\over (D-3) (D-2) } \right ) ^\h
\label{qfsix}
\ee
so it diverges in the Rindler limit $r_0\r\infty$. In other words, the location $s_h$ of the horizon depends on the total extent of the sheet in the directions $x_i$, and not just on the local surface energy density $\sigma$. (The extent of the sheet is $\Delta x_i\sim r_0$.)

In  section \ref{sec:6.2}, we had considered an analogy where the infalling object transitioned to fuzzballs at a location where the `height of the wave became comparable to the depth of the lake'.  In this language we can interpret (\ref{qfsix}) as follows. In general  the height of the incoming wave at a location $x_i$ is determined not just by the density $\sigma$ at the location $x_i$, but also by the value of $\sigma$ at neighboring values of $x_i$. If the sheet is very large, then the location $s_h$ will be determined by the entire extent of the sheet, and not just by its local energy density.

To study this in more detail, we now consider the infall of a compact object, where we can in fact take a Rindler limit. We will then see how to compose the effect of a distribution of such objects and recover the result (\ref{qfsix}) for the shell.

\subsection{\label{sec:8.3}Compact infalling objects}

Consider an infalling object with energy $E$. When this object gets absorbed by the black hole, the entropy of the hole increases by
\be
\Delta S \approx {E\over T}
\ee
In \cite{mtflaw} a rough criterion was proposed for when the tunneling into fuzzballs would commence. Suppose the infalling object is at a distance $s$ from the  horizon. Consider a hemispherical surface of radius $\sim s$ which represents the deformation of the horizon in the process of absorbing  the infalling object. The area of this surface is \be
\Delta A \sim s^{D-2}
\ee
Then it was conjectured that the tunneling into fuzzball will take place when the infalling object reaches a distance $s$ from the horizon where
\be
{\Delta A \over G} \sim \Delta S
\ee
which gives
\be
s\sim \left ( {E\over T} \right ) ^{1\over D-2} l_p
\label{qd}
\ee
It is useful to think of the energy at infinity $E$ in units of the energy $\sim T$ of the Hawking quanta at infinity
\be
n={E\over T}
\label{qn}
\ee
As the object falls in, the energy gets blueshifted to higher values in a local frame like that used in (\ref{qframe}), but $T$ gets blueshifted by the same factor, so that we always have $E_{local}/T_{local} = n$. In terms of $n$, the scale (\ref{qd}) is
\be
s\sim n^{1\over D-2}
\label{qscale}
\ee

In the above estimate we assumed that all length scales were $\sim s$, but  we can consider objects that are very compact, with a size $d\ll s$. In this case the location of the event horizon was computed in \cite{virmani}. For us the more relevant location is that of the apparent horizon, but for our present estimates we will assume that they are given by a scale that is at least qualitatively similar.   Consider the metric (\ref{qrindler}) describing the near horizon region.  We define Minkowski coordinates through
\be
T=s\sinh\tau, ~~~Z=s\cosh\tau
\ee
Consider a particle with energy given by $n$ as in (\ref{qn}). Let $\hat E$ be the energy of this particle when it is at a distance $s$ from the horizon,  as measured in a local Lorentz frame with axes along $T, Z$. At this location the local temperature is 
\be
T_{local}={1\over 2\pi s}
\ee
so 
\be
\hat E = {n\over 2\pi s}
\ee
 A pointlike object moving with the speed of light with this energy generates an Aichelberg-Sexl shock wave of the form
\be
ds^2 =-dUdV + \hat E\Phi(|x|) \delta (U-U_0)dU^2 + dx_idx_i
\label{qphi}
\ee
where $U=T+Z, ~V=T-Z$ and
\be
\Phi={c_D \over (D-4) |x|^{D-4}}
\ee
with
\be
c_D={16\pi G\over \Omega_{D-3}}
\ee
(We have restricted to $D>4$ for simplicity; the case $D=4$ gives a log in place of the power law in $|x|$.)

The future horizon is at $V=0$. Consider an outgoing null ray $V=V_0$ outside the horizon. Let us assume that this ray meets the shock when the shock is at the above selected distance $s$ from the horizon. In the process of passing through the shock the outgoing null ray will get `pushed' towards the horizon by an  amount
\be
\Delta V = \hat E \Phi (|x|)
\label{qshift}
\ee
where $|x|$ is the distance between the outgoing ray and the infalling particle as measured in the plane of the shock.\footnote{We have assumed that the outgoing ray emerges from the shock in a direction which is radially outwards (i..e travelling in the $Z$ direction); such rays will escape the pull of the hole more easily than rays directed at an angle. Thus we should consider radially outgoing rays  when locating the new horizon.} In this process we have $\Delta U=0$, which gives $\Delta T=-\Delta Z$. Thus
\be
\Delta V = -2\Delta Z
\label{qz}
\ee
If $\Delta Z=-s$, the outgoing null ray will get pushed into the horizon, and thus not emerge to infinity. Thus we  set the needed shift to $\Delta V=2s$, and equate this shift to the expression  (\ref{qshift}):
\be
2s= \hat E \Phi (|x|)=\left (  {n\over 2\pi s} \right ) \left ( {c_D \over (D-4) |x|^{D-4}}\right )  
\ee
Thus when the infalling particle is at a distance $s$ from the horizon, its shock front has trapped all outgoing null rays which are between  the shock and the horizon, as long as the value of $|x|$ for the null ray  is less than the one given by solving the above relation
\be
|x|_{max}= \left ({n c_D\over 4\pi s^2} \right ) ^{1\over D-4}
\label{qxmax}
\ee
To summarize, when the infalling particle is at a distance $s$ from the horizon, the excitation  in the region $|x|\lesssim |x|_{max}$ at the radial position $s$ has reached the `bottom of the sea' in the schematic picture of fig.\ref{fig3}, and we will have a tunneling into fuzzballs in this region. 

If we set all scales to be comparable; i.e., $|x|_{max}\sim s$ in (\ref{qxmax}), then we recover (\ref{qscale}). 

\subsection{Infall of a mass distribution}

Let us now consider more general distributions of infalling matter. First, let us relate the location of the horizon (\ref{qffive}) we found for the spherical shell to the approach we used for the infalling particle. Consider again a infalling shell with mass $m$; at a distance $s_1$ from the horizon its surface energy density is given by (\ref{qframe}). Suppose this shell falls inwards to a location $s<s_1$. The surface energy density is now
\be
\sigma = \sigma_1 \left ( { s1\over s} \right ) 
\ee
The quantity $\hat E \Phi$ in (\ref{qphi}) is give by integrating over the shell to find the potential. Thus at $x=0$ we would get
\be
\hat E \Phi(x=0) \sim \int d^{D-2 } x' G\sigma\,  {1\over |x'|^{D-4}}
\ee
For an infinite plane this integral would diverge, but we should cut off the integral at $|x|\sim r_0$. Thus we set
\be
\int d^{D-2 } x {1\over |x|^{D-4}}\sim r_0^2
\ee
We then get for the shift $\Delta V  $
\be
\Delta V  \sim G\, \sigma_1 \left ( { s_1\over s} \right ) r_0^2
\ee
Setting $\Delta V = 2 s$ as in (\ref{qz}) we find that a horizon will appear when
\be
s\sim G \sigma_1 \left ( { s_1\over s} \right ) r_0^2
\ee
This gives
\be
s\sim r_0 \left ( G\sigma_1 s_1\right ) ^\h 
\ee
which agrees with (\ref{qfsix}).  

For more general distributions of infalling matter, we can similarly compute the potential $\Phi$ created by the shock  waves in the near horizon region, and compute the shift $\Delta V$ due to these shocks. If this shift pushes a geodesic back by an amount that would take it inside the surface of the existing fuzzball, then we can say that this geodesic is `trapped', and use this to fact to get as estimate of when the tunneling into fuzzballs should commence.

\section{\label{sec:4}The causality problem and the firewall argument}

To see the power of the causality constraint, we now show that it creates a conflict between two of the assumptions made in the firewall argument.

Hawking's computation showed that if we have the vacuum state at the horizon, then we will have a monotonically increasing entanglement. This leading order computation was made rigorous using a bit model in \cite{cern}, where it was shown to be robust against small corrections to the evaporation process.  This converts the Hawking argument into a `theorem'. We can state the theorem in an exactly equivalent way as follows. Suppose we assume 

\mm

{\bf Ass:1} The information in the hole is radiated out the same way as by any other black body; i..e, there is no monotonic rise in entanglement.

\mm

Then the Hawking theorem says that the horizon cannot be a vacuum region. 

\mm

AMPS \cite{amps} sought to make this result stronger by adding an extra assumption

\mm

{\bf Ass:2} Let the region $r>r_h+l_p$ (i.e., the region outside the stretched horizon) be described by `effective field theory';  i.e., the physics outside the hole is `normal physics'. In particular, if a shell is approaching the stretched horizon at the speed of light, then, by causality, the stretched horizon cannot respond in any way until the shell actually reaches the stretched horizon. 

\mm

Under these assumptions, AMPS argued that an infalling object will encounter radiation quanta of increasingly high energy $E_{rad}$ as it approaches the horizon, with  $E_{rad}$ reaching planck scale at  the stretched horizon. Thus not only is the region near the horizon not a vacuum, it is a `firewall' for any object that tries to enter the hole. 

The intuition behind the firewall argument is simple. In Hawking's pair creation from the vacuum, the particles do not actually materialize until they are well separated from the horizon; the region around the horizon remains a vacuum. Thus any actual particles (i.e. those that can be interacted with)  are always long wavelength ($\lambda \sim r_h$) quanta. But if the radiation was emerging from a hot surface placed at the stretched horizon, then one can follow these quanta back to a location close to the stretched horizon, where they will be blueshifted to high energies. They will still be real particles however, and can interact with and burn an infalling object. 

But we find that there is a problem with this argument, since Assumptions Ass:1 and Ass:2 are in conflict with each other due of the causality problem. We can see this  as follows:

\mm

(a) Consider a black hole of mass $M$. The stretched horizon is at
\be
r_s=2GM+l_p
\ee
By assumption Ass:2,   the region $r>r_s$ has `normal physics', given by usual effective field theory.

\mm

(b) Now consider a shell of radially ingoing gravitons, carrying a total energy $\Delta M$
 Since this shell moves at the speed of light, it continues to move inwards all the way to $r=r_s$, with a dynamics governed just  by effective field theory (again by assumption Ass:2).
 
 \mm
 
  (c) The total mass of the black hole and the shell is  $M+\Delta M$, which corresponds to a horizon at the location,
\be
r=2G(M+\Delta M)
\label{qttone}
\ee
From (a) we see that the shell must pass without drama through the location (\ref{qttone}). But then  the information in the shell is trapped inside its own horizon, and cannot reach infinity unless we have a violation of causality.\footnote{One should note that the shell can interact with any Hawking radiation quanta that are emerging from the stretched horizon. But the energy of these quanta drops sharply as they recede from the horizon. The wavelength at a distance $s$ from the horizon is $\lambda \sim s$, so that at 1 mm from the horizon the temperature as already lower than the microwave background of outer space. Thus we can easily take $\Delta M$ large enough so that the shell has a negligible interaction with this radiation at the point where it passes through its horizon.} 

\mm

(d)  But we cannot violate causality in the region $r>r_s$, since this region is assumed to be described by effective field theory. Thus we find that the information in the shell {\it cannot} be  radiated to infinity, in violation of assumption Ass:1.\footnote{We can of course let the information be trapped in a remnant, and then perhaps leak out very slowly over a timescale much longer than Hawking evaporation time. But this is not what was assumed in Ass:1 -- this assumption was  really asking for the hole to radiate like a normal body and send its information out in the radiation.} 

\mm

Note that if we are willing to violate causality, then there is no information puzzle in the first place; we can always say that some mechanism takes the information from the singularity and puts it outside the hole. Thus we see that causality creates a conflict between the assumptions Ass:1 and Ass:2  used in the firewall argument; in consequence we cannot argue that black hole horizons must act like firewalls. 

Let us now see how the assumptions of the firewall argument differ from the situation in the fuzzball paradigm. 
The crucial point is that in the fuzzball paradigm, at any point $r>r_s$ outside the  hole,   we have `normal physics' (effective field theory) {\it only for infalling objects up to a certain energy $E(r)$}; objects with energy $E<E(r)$ will travel through the location $r$ without  any significant departure from semiclassical evolution, while objects with energy $E>E(r)$ will not. Recalling the infall picture of section \ref{sec:7.3}, we see that the transition to fuzzballs prevents the trapping inside the horizon. By 
contrast, the assumptions of the firewall argument force the trapping of a massive shell inside its own horizon, and then we have a problem with causality.

\section{\label{sec:4c} Fuzzball complementarity}

In the above sections we have seen how the fuzzball paradigm can solve problems (A),(B) while maintaining the requirement of causality in our theory of gravity. In this section we will recall the conjecture of fuzzball complementarity, which addresses a somewhat different question in black hole dynamics: the {\it infall problem}. After recalling this conjecture we will note that while causality is certainly not sufficient for the  conjecture to be true, it {\it is} a necessary condition for such a conjecture to be possible. 

\subsection{\label{sec:9.1}The infall problem}

Besides the questions (A) and (B) stated in section \ref{sec:1}, we can ask another question:

\mm

(C) What does an infalling observer feel as he reaches the horizon of a black hole?

\mm

In the traditional semi-classical picture of a black hole he would pass through the horizon without noticing anything special.  But if the horizon is a vacuum region
then we have the problems (A),(B). If we have a  nontrivial
structure at the horizon, then we may be able to resolve (B), but it seems that the
observer must interact with this structure and thus not
feel that he is harmlessly passing into the black hole
interior.

We may call (C) the `infall'. question. Note that this
question is not on the same footing as questions (A),(B),
in the sense that it does not relate to any fundamental
problem with our theory of gravity. If the observer feels
something nontrivial at the horizon, then we say that the
full quantum gravity theory implies for him an experience
different from the traditional semiclassical expectation.
But this difference is not in conflict with some basic law of physics; besides,
we have to find a violation of the semiclassical approximation
anyway to resolve the problems (A),(B). On the
other hand the problems (A),(B) are fundamental difficulties:
  (A) conflicts
 with causality, while (B) conflicts
with the basic requirements of linear quantum mechanics.

In spite of the fact that there is no fundamental requirement on the nature of infall, it would be more satisfying if the classical intuition of `free fall through the horizon' were preserved in some way by the full quantum theory. This desire to preserve free infall gave rise to the notion of `complementarity'. The term    arose in the work of 't Hooft \cite{thooft} and the idea was formulated in detail by Susskind and others \cite{susskind}.  In the latter approach one postulates that `new physics' arises when a black hole forms: the information in an infalling object can be {\it duplicated}. An observer outside the hole sees the information returned from the horizon to infinity, while an observer falling in with the object sees the information being carried into the hole. Normally we cannot duplicate information (`no cloning') because such a duplication process conflicts with the linearity of quantum mechanics. But observers inside the hole cannot communicate with the outside, so one cannot easily compare the duplicate copies inside and outside the hole; it was argued that this fact allows us to bypass the `no cloning' theorem when a black hole forms. 

There are several immediate difficulties with such a complementarity proposal. The principal one stems from the fact that the classical picture of gravitational collapse can be studied in a `good slicing' where nothing special happens at the horizon. What then triggers `new physics' when a horizon forms? What mechanism reflects information back to infinity for an outside observer? It was argued that the outside observer must use Schwarzschild coordinates at the horizon, and quantum field have large fluctuations at the horizon in these coordinates. But such fluctuations would appear to a {\it coordinate} artifact, and so it is not clear how they could lead to a reflection of information from the horizon. Lastly, if the good slicing picture were true, then how do we stop the production of entangled pairs and the corresponding problem (B)?

We will call the above proposal of complementarity as `traditional complementarity'. This will serve to distinguish it from a conjecture about infall that we can make in the fuzzball paradigm; the latter will be termed `fuzzball complementarity'.

\begin{figure*}
 \includegraphics[scale=.22] {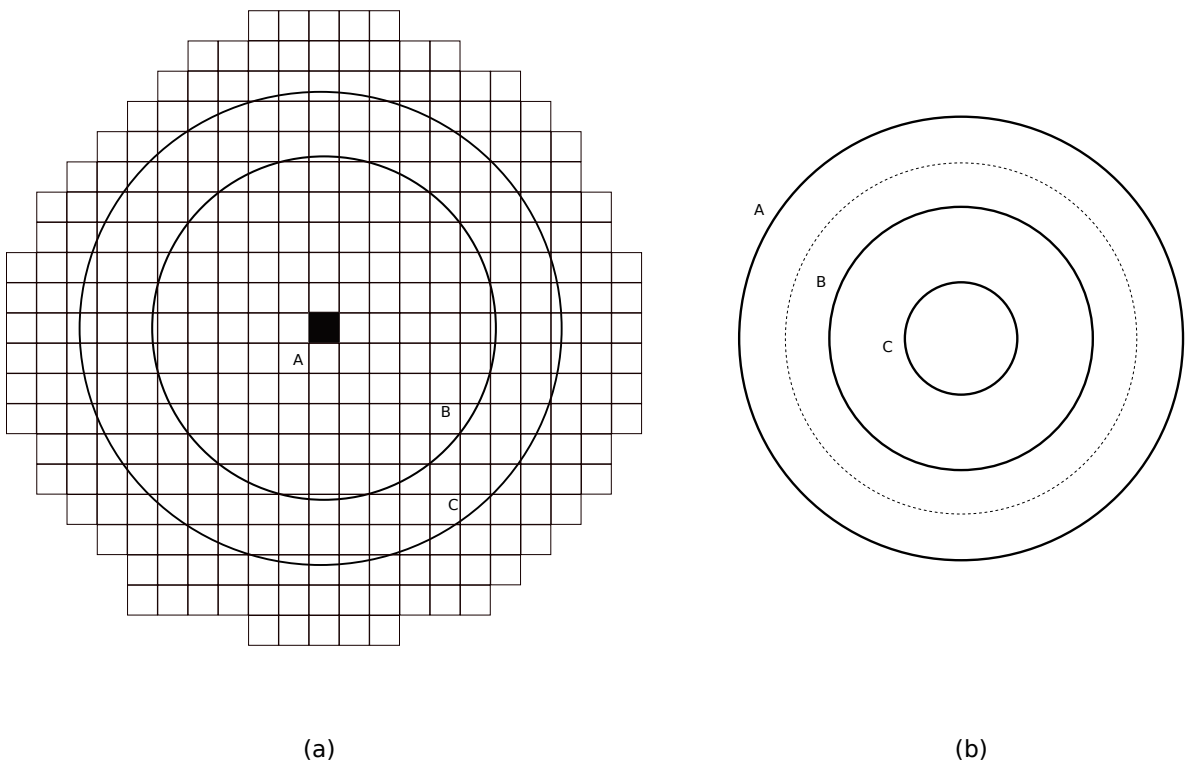}
\caption{The idea of fuzzball complementarity. (a) The configurations in superspace. The central dark square represents an incoming shell of mass $M$, when it is travelling semiclassically at $r>2GM$. After the shell reaches $r\approx 2GM$, its wavefunction spreads in superspace, since the shell tunnels to a linear combination of fuzzballs.  This wavefunction in superspace reaches the circle B and at a later time, the circle C. (b) The emergence of  an effective black hole interior. The shell travels semiclassically at position A, where it is outside its horizon (the dotted circle). The locations B and C do not actually exist in the spacetime, but the progression of the wavefunction indicated in (a) can be  depicted by using an effective  black hole interior where the shell continues to move to $r<2GM$.} \label{fig10}
\end{figure*}

\subsection{\label{sec:9.2}The conjecture of fuzzball complementarity}

With fuzzballs we find a rather different situation from the one which led to the ideas of traditional complementarity. With traditional complementarity, the goal was to reconcile two facts: (i)  the horizon is in the local vacuum state $|0\rangle$ for an infalling observer and (ii)  information should escape from this horizon in Hawking radiation. With fuzzballs, the semiclassical approximation is violated by `entropy enhanced tunneling', and $r\approx r_h$ is {\it not} a vacuum region.  Further, in the fuzzball paradigm we require no `new physics' in the presence of a black hole: all dynamics of the hole must follow from just the usual rules of string theory which is based on linear quantum mechanics. 

In this situation it may appear that an infalling observer must fell that he `crashes and burns' at the surface of the fuzzball. Indeed the firewall argument \cite{amps} attempted to argue that any object like a fuzzball that radiated information from its surface will have to necessarily behave like a firewall for infalling observers. We have seen that the assumptions used in the firewall argument are in conflict with one another if we assume that causality holds in the underlying theory. In the picture of fuzzball formation presented above, we do maintain causality, so the firewall argument cannot really  apply  to this picture. But one may still wonder if some modification of the firewall argument could rule out any feeling of  smooth infall in the fuzzball paradigm. 

 In a series of papers \cite{plumberg,beyond,mtflaw}, a scenario was developed which allowed an infalling observer to  feel no violent impact when he reaches $r\approx r_h$, while information was still unitarily radiated from the surface of the fuzzball. The key point was that this feeling of `free infall' was limited to a subclass of observers: those who fall in freely from afar with an energy $E\gg T$, where $T$ is the temperature of the hole. Since $T\sim {1\over r_h}$ is very small for a large hole, we see that objects with a given energy $E$ will feel very little unusual behavior at the horizon as the mass of the hole is taken to infinity. 
 
 In more detail, the conjecture of fuzzball complementarity is as follows:
 
 \mm
 
 (1) Consider the gravitational collapse of a shell  depicted in fig.\ref{fig10}(b). At position A, the shell is far from its horizon, and its evolution is given by semiclassical gravity. 
 
 In fig.\ref{fig10}(a), we depict {\it superspace}, the space of all solutions of our quantum gravity theory. The wavefunction of the shell at position A is depicted schematically by the square at the center of fig.\ref{fig10}(a). 
 
 \mm
 
 (2) As the shell reaches its horizon $r\approx r_h$, it will tunnel into a linear combination of fuzzballs  as discussed in section \ref{sec:3}.  The evolution of the full quantum state is depicted by an approximately spherical wavefront in superspace, with the radius of this wavefront moving to larger values as the evolution progresses: thus the full state of the shell evolves to the wavefront B and then to the wavefront C in fig.\ref{fig10}(a).
 
 \mm
 
 (3) The conjecture of fuzzball complementarity says that this evolution ${\rm B}\r {\rm C}\r \dots$ can be approximately mapped to the infall of a shell in the traditional picture of the black hole; i.e., the infall depicted by the locations ${\rm B}\r {\rm C}\r \dots $ in fig.\ref{fig10}(b). Thus the actual exact wavefunctional in the full gravity theory never forms a horizon, but the  evolution in {\it superspace}    can be mathematically mapped to a picture where the shell is allowed to progress into the interior of an {\it effective} black hole geometry. This picture is approximate, becoming more and more accurate as we let $E/T\r \infty$. (Here $E$ is the energy of the shell, and $T$ is the temperature of the hole, which in the present case is $T\sim 1/E$.)

 \subsection{\label{sec:9.3}A partial analogy}
 
 \begin{figure}
 \includegraphics[scale=.62] {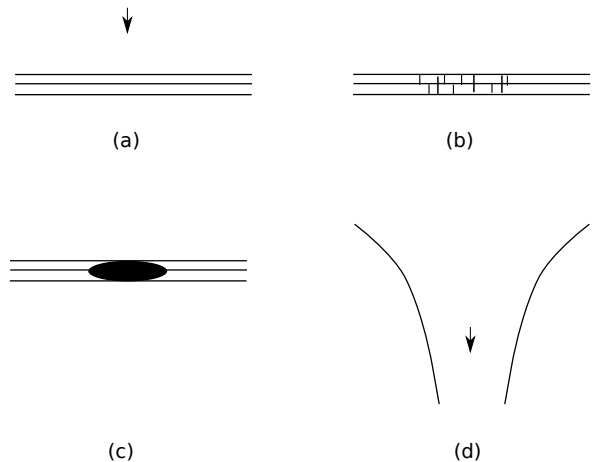}
\caption{\label{figadscausality} (a) A graviton is incident on a stack of D3 branes. (b) When the graviton hits the branes, it creates open strings, so it may appear that the graviton has been `destroyed'. (c) The open strings are, however, in a definite coherent state; we denote this by a blob which expands along the surface of the branes while maintaining its coherent structure. (d) The expansion of the blog along the D3 branes can be mathematically mapped to the progression of the graviton into an AdS space; in this way we see that the  graviton is not `destroyed'. }
\end{figure}

 To understand the  fuzzball complementarity conjecture better, we note that something similar happens in the case of  AdS/CFT duality. It will be important to note both the similarities and the differences between the two situations.
 
 Consider a stack of $N$ coincident D3-branes, depicted in fig.\ref{figadscausality}(a). Let a graviton be incident on this stack. When the closed string representing the graviton reaches the branes, it gets converted into a collection of open strings with endpoints on the branes (fig.\ref{figadscausality}(b)). It would therefore seem that the graviton has been destroyed in the process of being absorbed by the branes.
 
 But we can replace the branes by a smooth AdS space as depicted in fig.\ref{figadscausality}(d), and in this description the absorption process is replaced by an uneventful passage of the graviton into the AdS throat. In this description it does seems that the graviton has {\it not} been destroyed. How does this description square with the earlier one in terms of open strings?   
 
 The explanation lies in the strongly coupled nature of the gauge theory describing the open strings. While the pictorial description in terms of open strings suggests that a  graviton  can transition into a large number of open string configurations, the strong interactions between these open strings forces most of these states to be lifted to high energies. This leaves  relatively few low energy states that are relevant for the process in question, and these states have a definite coherent structure in terms of their distribution of open strings. Thus we depict these low lying states by a coherent blob on the D-branes (rather than a gas of open strings); this is depicted in fig.\ref{figadscausality}(c). As time passes, this blob expands on the surface of the D-branes, while maintaining its  coherent internal structure. This evolution of the blob along   the surface of the branes may be mapped, mathematically,  to the simple evolution of the graviton into an AdS interior. This is the way AdS space emerges from strongly coupled gauge theory. 
 
 Remarkably, there {\it do} exist a large number states in the gauge theory that are not lifted to high energies: these are the states corresponding a black hole and they have a high degeneracy. But these states  are not easily accessible when we start with simple configurations like those corresponding to a few high energy gravitons. The blobs corresponding to the high energy gravitons first spread over all the available D-branes, and then the state  evolves to a quark-gluon plasma phase;  such plasma states are the ones that are expected to yield the entropy of a black hole.

Now consider the collapse of the shell depicted in fig.\ref{fig10}:

\mm

(i)  When the shell is outside its horizon (position A), the situation is analogous to the graviton being away from the D3 branes.

\mm

(ii) When the shell reaches  $r\approx r_h$ it tunnels to fuzzballs, and the further evolution of the quantum gravity state must be described in superspace, the space of all complicated gravity configurations. This evolution is depicted by wavefronts B, C in fig.\ref{fig10}(a).   The expanding wavefront in superspace is analogous to the expanding blob made of open strings on the D3 branes.

\mm

(iii) It may appear that there are many directions in superspace, and so the evolution of the wavefunctional could be a very complicated spread in all these directions. But there is a strong coupling between neighbouring configurations in superspace; we can think of this coupling as arising from   a large transition amplitude between neighbouring  fuzzball configurations. This coupling lifts a large number of supersapce wavefunctionals to high energy, leaving very few low energy wavefunctionals that are relevant to the process in question. As a toy model, we can think of a 2-level system with Hamiltonian
\be
\hat H = \hat H_0+\lambda \hat H_1 =  \begin{pmatrix} E & 0 \cr 0 & E \cr  \end {pmatrix} + \mu \begin{pmatrix}  0 & 1 \cr 1 & 0 \cr \end{pmatrix}
\ee
The transition amplitude $\mu $ lifts the degeneracy, giving a low energy eigenvector $(1, 1)$ with eigenvalue $E-\mu$ and a high energy vector $(1, -1)$ with energy $E+\mu$. In a similar way, the low energy configurations in superspace that are easily accessed from the initial shell state $|S\rangle$ are depicted as nearly spherical wavefronts in superspace where the wavefunction has spread almost uniformly over a large number of fuzzball configurations. 

\mm

(iv) The evolution of the wavefronts in superspace can be mathematically mapped onto the motion of a shell that is progressing past $r\approx r_h$ into the interior of a traditional hole.  As we will discuss below,  it is very important that this mapping is approximate rather than exact, with the approximation improving as we let $E/T \r\infty$. This map is analogous to the map in AdS/CFT duality where the expanding blob in the configuration space of open strings is mapped to a graviton progressing deeper into an AdS spacetime. 

\mm

(v) There exist a large number of wavefunctionals in superspace that are {\it not} lifted to high energies, but these are not easily accessible when we start with the simple initial shell state $|S\rangle$.  These wavefunctionals correspond to  generic black hole states, and account for the Bekenstein entropy $S_{bek}$. After the wavefunctional has moved through the stages  ${\rm B}\r {\rm C}\r \dots$ and  in the approximate description of fig.\ref{fig10}(b) the shell has reached near $r\approx 0$, the wavefunctional in superspace starts to spread in the direction of these generic states of superspace, and this is the process of relaxation towards the generic fuzzball states which are characterized by the entropy $S_{bek}$. 

\mm

It would seem from the above discussion that the conjecture of fuzzball complementarity is very similar to the conjecture of AdS/CFT duality: just as the AdS space emerges from the dynamics of collective excitations on  D-branes, the interior of the hole can emerge from the collective dynamics of fuzzballs. But as we will now see, there is a crucial difference:  fuzzball complementarity only requires an effective black hole interior to emerge  when the infalling shell has energy $E\gg T$.

 \subsection{\label{sec:9.4}The condition $E\gg T$}

 \begin{figure}
 \includegraphics[scale=.32] {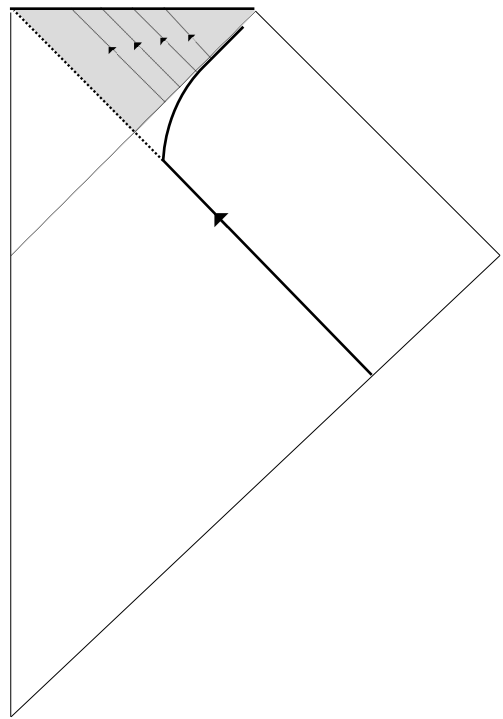}
\caption{\label{fig8causality} A Penrose diagram illustrating the conjecture of fuzzball complementarity. The thick left moving line represents a shell that is collapsing in flat spacetime. Just before the shell reaches the horizon, it tunnels to fuzzballs; the fuzzball surface is depicted by the thick line just outside the position where the horizon would have been. The shaded region of spacetime does not exist; it can however be added as an approximate mathematical description of high energy infalling objects. These objects create an expanding wavefront in superspace, and the progression of this wavefront can be mathematically mapped to infall in the shaded region. The left moving arrows in the shaded region tell us that this region is an effective spacetime only for high energy left moving objects; it is not an effective spacetime for Hawking modes, which are right moving quanta.}
\end{figure}

In AdS/CFT the gauge theory dynamics of D-branes gives rise to AdS spacetime. Suppose the dynamics of fuzzballs really gave rise to the traditional geometry of a black hole. Then we would face an immediate problem: won't the low energy dynamics around  the horizon  produce entangled pairs, just as in Hawking's original computation?  If we get such pairs, then we will be back in problem (B), the problem of monotonically growing entanglement. 

It would also seem that we cannot avoid this problem by asking that the traditional geometry of the horizon emerge only approximately. The small corrections theorem \cite{cern} says that the entanglement of the produced pairs is robust: small corrections to the state of the entangled pairs cannot change the conclusion that the entanglement keeps rising monotonically. Should we therefore conclude that fuzzball complementarity is not possible; i.e., any such conjecture would destroy the resolution we have found to puzzles (A) (B) through the fuzzball paradigm?

The key to the conjecture of fuzzball complementarity is that the effective black hole interior only emerges for  objects that are {\it infalling} with high energy onto the location $r\approx r_h$. Consider a black hole of mass $M$; this would be in a generic fuzzball configuration with radius $r_h\approx 2GM+\epsilon$. Let a shell of energy $E\gg T \sim 1/r_h$ fall onto this hole. We have the following situation:

\mm

(i) The low energy outgoing quanta with energy $E\sim T$ have no description in terms of semiclassical modes on a traditional black hole background. These modes are emitted from the fuzzball surface and carry the information of the fuzzball, just as photons emitted from a piece of coal carry information about the coal. 

\mm

(ii) {\it Infalling} objects with energy $E \gg T$ have an approximate description in terms of motion in an effective black hole interior; this description is accurate only for a brief period of infall, during which the object travels in an effective geometry between the locations $r\approx r_h$ and $r\approx 0$. It is these modes for which the wavefunctional in superspace spreads in the manner depicted in fig.\ref{fig10}(a). 

\mm

The reason for the difference between (i) and (ii) has to do with how many states can be accessed with the given energy; we will discuss this in more detail below. The Penrose diagram for illustrating the fuzzball complementarity conjecture was given in \cite{bitmodel}, and is reproduced in fig.\ref{fig8causality}. The actual smooth spacetime ends just outside $r\approx r_h$ at the fuzzball boundary. The region $0<r< r_h$ is depicted with left-directed arrows, to signify the fact that it arises as an effective description only for high energy {\it infalling} objects. 

Let us now come to the difference between cases (i) and (ii) above, and in particular how differentiating between these two is important in bypassing a firewall type argument.  In the firewall argument one focuses on the entanglement of quanta emitted from $r\approx r_h$ which will end up as Hawking radiation quanta at infinity. (These are the quanta that we term $E\sim T$ quanta, whatever be their position; their energy in a local frame near the horizon will of course be blueshifted to a larger value.) But these quanta are {\it not} those for which we ask for  any complementary description in fuzzball complementarity. Instead we focus on quanta which start at infinity with $E\gg T$ and fall in. Such quanta create a large number of {\it new} degrees of freedom when they approach the fuzzball surface, and these new degrees of freedom are not entangled with anything. If we start with a hole of mass $M$ and add an energy $E$, the total number of final states 
is related to the number of initial states as
\bea
{{\cal N}(M+E)\over {\cal N}(M)} &=& {Exp[{S_{bek}(M+E)}]\over Exp[S_{bek}(M)]}\nn
&\approx& Exp[\Delta S_{bek}]\approx Exp[E/T]\gg 1
\label{qwone}
\eea
Thus a high energy infalling quantum creates a large number of new degrees of freedom, and it is the evolution of these (unentangled) new degrees of freedom that is described by the conjecture of fuzzball complementarity. 

We can now see the difference between the cases (i) and (ii) above. In case (ii) we have $E\gg T$, and the  large number of new degrees of freedom (\ref{qwone}) have very little to do with the initial state of the fuzzball before the new energy $E$ was added. Thus these evolution of these new degrees of freedom can have a universal form given by the wavefronts in fig.\ref{fig10}(a). But in case (i) the infalling quantum adds only $\sim 1$ new bits to the fuzzball. The quantum reaches within $\sim l_p$ of the fuzzball surface before creating the new bit, and so its evolution is strongly influenced by the initial state of the fuzzball. Thus there is no approximation of `effectively smooth infall' that is available for such $E\sim T$ quanta.

In \cite{bitmodel} a bit model was presented where information escaped in low energy  ($E\sim T$) modes while the high energy ($E\gg T$) infalling modes had an effective evolution mimicking infall into a black hole interior. This model therefore gives an explicit counter to the firewall argument.

 \subsection{\label{sec:9.5}Causality and fuzzball complementarity}

Finally let us come to the relation between fuzzball complementarity and the subject of the present paper: causality. 

The conjecture of fuzzball complementarity uses the fact that an infalling object transitions to fuzzballs just before the place where it would have created the horizon. Having a transition at this location (or earlier) is required by causality, as we have noted.  The spread of the wavefunctional in superspace then gives an effective evolution that can be interpreted as infall into a black hole interior. 

Suppose on the other hand we had the set-up assumed in the firewall argument of \cite{amps}. Let the infalling object be composed of massless radially infalling quanta. In this case the assumptions of the firewall argument  would demand that the object approach the horizon of the existing hole without any novel quantum gravity effects being relevant. The infalling quanta would however interact with the radiation being emitted from $r\approx r_h$. This interaction would destroy the infalling object, and then we cannot hope for any complementary description in which one would see free infall through the horizon. 

To make this concrete, let us outline argument of \cite{amps} in the present context. Let the hole be past its halfway point of evaporation. A radiation quantum $h$ being emitted from the hole has a state which is close to maximally entangled with states $|out\rangle_i$ at infinity, so that the relevant part of the quantum state is
\be
{1\over \sqrt{2} }  \Big(\,  |+\rangle_h |out\rangle_1 + |-\rangle_h|out\rangle_2 \, \Big ) 
\ee
where $|\pm\rangle_h$ are the spin states for $h$. Now consider the infalling object to be a quantum in a state $|q\rangle$. This quantum can scatter off the two spin states of $h$ in different ways; for example
\be
|q\rangle |+\rangle_h \r |q_1\rangle |+\rangle_h, ~~~ |q\rangle |-\rangle_h \r |q_2\rangle |-\rangle_h
\ee
where $|q_1\rangle, |q_2\rangle$ are two orthogonal states of $q$. Since the energy of the radiation quantum $h$ goes to the planck scale as we follow it back to the stretched horizon, the probability of such a scattering is high if we are allowed to follow $q$ all the way to the stretched horizon.  After this scattering the overall state of $h, q$ and the radiation at infinity has the form
\be
|\Psi\rangle = {1\over \sqrt{2} } \Big  (\,  |+\rangle_h |out\rangle_1 |q_1\rangle + |-\rangle_h|out\rangle_2|q_2\rangle \,  \Big  ) 
\ee
Now we cannot get any complementary behavior for $q$. In obtaining complementarity we are allowed to make a change of basis,  and with this change of basis the evolution of $q$ must become one where $q$ did not change its state when it passes through $r\approx r_h$.  But should this change of basis map $|q\rangle \r |q_1\rangle$ or 
$|q\rangle \r |q_2\rangle$? Since we can make at most one of these choices, we see we cannot make a change of variables where  $q$ feels no change at the horizon. 

But as we have noted above, if we let $q$ reach the stretched horizon as required by the assumptions of the firewall argument then we have a problem with causality: $q$ is trapped inside its own horizon and cannot sent its information to infinity without information transfer outside the light cone. In the fuzzball paradigm the $q$ transitions to fuzzballs before reaching the stretched horizon. The distance from the stretched horizon where this transition occurs increases with $E/T$. Thus for large $E$  the radiation at the point of transition is  not strong enough to give any significant interaction between $h$ and  $q$ \cite{mtflaw}. The effective infall description is then encoded in the new degrees of freedom (counted as in (\ref{qwone})) which are created by the energy $E$, and the subsequent evolution in superspace of the fuzzball configuration (fig.\ref{fig10}) generated by these degrees of freedom. Thus we see that fuzzball complementarity evades the firewall argument by the same mechanism by which it avoids the causality problem.

\section{\label{sec:10} Discussion}

Let us  note the three different categories of solutions that have been proposed to deal with the information paradox:

\mm

(1) {\it Remnants:}  The data of the infalling shell (and the negative energy partners of Hawking radiation quanta) stay trapped in a planck sized object.

\mm

(2) {\it Fuzzballs:}  String theory yields states that do not collapse under their own gravity, and such states describe the microstates of black holes. Thus there is no horizon or singularity. In particular the absence of a vacuum region around a horizon implies that we do not get the creation of entangled pairs by Hawking's mechanism \cite{hawking}, and this resolves the entanglement problem (B). 

\mm

(3) {\it Wormholes:} The horizon is a vacuum region, but novel physics intervenes to resolve the information puzzle. The new physical concept is that the degrees of freedom near infinity are not distinct from the degrees of freedom inside the horizon \cite{cool, pr}. We may depict this identification of degrees of freedom by a set of thin wormholes that connect the interior of the hole to the radiation quanta at infinity. More generally, one could conjecture nonlocal effects over different length scales: Giddings \cite{giddings} has conjectured nolocality for low energy modes over scales $\sim GM$, while Hawking et. al \cite{hps} have conjectured that the degrees of freedom of the hole may live at infinity, being encoded in diffeomorphisms that do not vanish at the boundary of spacetime. 

\mm

We now summarize the various aspects of the fuzzball paradigm (2), and then note the relation between these aspects and the issue of causality that we have discussed in this paper. The fuzzball paradigm has four different results/conjectures, that are  loosely related  with each other:

\mm

(i) The actual fuzzball construction, which gives explicit examples    of black hole microstates in string theory with no horizon or singularity \cite{lm4,fuzzballs}.

\mm

(ii) The small corrections theorem, which says that we cannot remove the problem of growing entanglement by small corrections to the states of the created pairs \cite{cern}.

\mm

(iii) The conjecture that the semiclassical approximation is violated at the horizon scale by `entropy enhanced tunneling' into fuzzballs \cite{tunnel,kraus,puhm}. 

\mm

(iv) The conjecture of `fuzzball complementarity', which says that it may be possible to preserve a notion of approximate semiclassical infall for infalling observers that fall in freely from afar with $E\gg T$ \cite{plumberg,beyond,mtflaw}.

 \mm
 
From (ii) we see that we must have an order unity change to the evolution at the horizon. One may try to avoid this conclusion by having a  nonlocal identification of degrees of freedom of the kind proposed in the wormhole scenario. But if we have nonlocality, then we typically lose causality as well. The construction (i) is of course the central feature of the fuzzball paradigm, which makes the rest of the picture possible. In a theory which did not have fuzzballs, an infalling shell would pass smoothly through its horizon, and we would then need to violate causality if we wish to have its information emerge in Hawking radiation. The conjecture (iii)  allows the fuzzball effects to start at the location where a horizon would have appeared, and this is needed to prevent trapping inside the new horizon. Finally, the same effects that give causality also allow us to have the conjecture of fuzzball complementarity (iv), though this conjecture is not required by causality. 

The firewall argument \cite{amps} assumes that an infalling shell will see no novel effects of quantum gravity until it reaches $r\approx 2GM+l_p$; this is called `validity of effective field theory' outside the stretched horizon'. In this situation a shell of mass $\Delta M$ will be trapped inside its own horizon at $r=2G(M+\Delta M)$, and we will have to violate causality if we require that its information emerge in Hawking radiation. We have noted that this sets up a contradiction between two of the postulates assumed in the firewall argument. (If we do not assume that effective field theory is valid outside the stretched horizon then we cannot prove that there must be a firewall; in fact in this situation we can construct an explicit bit model \cite{bitmodel} which gives `fuzzball complementarity' for $E\gg T$ infalling objects.)

From the perspective of our picture, the problem with  the firewall approach arises from asking the question: is effective field theory valid outside the stretched horizon or not?  In the actual situation conjectured here, effective field theory is valid at a given position $r>r_b$ {\it for objects upto a certain energy $E$, and not for objects of higher energy}. With this picture we resolve all the problems with the quantum theory of black holes.   

Finally, one may ask if it is essential to require causality in a theory of quantum gravity. One may argue that quantum fluctuations of the metric will lead to a fluctuation of light cones, so that no strict causality is possible. But perturbative diagrams in general relativity and string theory preserve causality, and the nonperturbative aspects of string theory have shown no violation of causality either. Even the very nonperturbative gravity process of bubble nucleation in a false vacuum respects causality: the nucleated universe expands at a speed less than the speed of light as seen from the both the true and the false vacuum regions. Note also that some fundamental approaches to developing a theory of quantum gravity have the  
notion of causality built in from the start \cite{cdt,sorkin}.

To understand how there can be some notion of causality in a theory with fluctuating metrics,  consider a maximally symmetric space like Minkowski spacetime. This symmetry group is defined using coordinates $x^\mu$ and a fiducial metric $\eta_{\mu\nu}$. The fluctuations around this fiducial metric  are not completely arbitrary: they must satisfy the requirement that the full wavefunctional preserve the Poincare symmetry group.  We can then define causality using the light cones of the fiducial metric $\eta_{\mu\nu}$; i.e., ask that commutators of local field operators vanish strictly outside the light cones defined by $\eta_{\mu\nu}$. 

The Schwarzschild metric   is not a maximally symmetric space, but the curvature away from the singularity is small, and so any ambiguities in defining causality in the region  $r\gtrsim 2GM$ should also be small. It is therefore unlikely that the information puzzle is resolved by effects that violate causality. If we assume that causality holds, then we have argued that the only way to get information to emerge in the Hawking radiation is to have an alteration of quantum fluctuations in the region outside the fuzzball; i.e., replace the Rindler region of the traditional black hole by pseudo-Rindler space.

\section*{Acknowledgements}

I am grateful to Borun Chowdhury,  Sumit Das, A. Jevicki, Oleg Lunin, Emil Martinec,  David Turton and Amitabh Virmani for helpful discussions. This work is supported in part by DOE grant de-sc0011726, and by a grant from the FQXi foundation.


\begin{thebibliography}{99}

 \bibitem{hawking}
  S.~W.~Hawking,
  Commun.\ Math.\ Phys.\  {\bf 43}, 199 (1975)
  [Erratum-ibid.\  {\bf 46}, 206 (1976)];
  S.~W.~Hawking,
  Phys.\ Rev.\  D {\bf 14}, 2460 (1976).
  

  
  




\bibitem{cern}
  S.~D.~Mathur,
  Class.\ Quant.\ Grav.\  {\bf 26}, 224001 (2009)
  [arXiv:0909.1038 [hep-th]].
  

  \bibitem{adscft}
  J.~M.~Maldacena,
  Adv.\ Theor.\ Math.\ Phys.\  {\bf 2}, 231 (1998)
  [Int.\ J.\ Theor.\ Phys.\  {\bf 38}, 1113 (1999)]
  [arXiv:hep-th/9711200];
  E.~Witten,
  Adv.\ Theor.\ Math.\ Phys.\  {\bf 2}, 253 (1998)
  [arXiv:hep-th/9802150];
  S.~S.~Gubser, I.~R.~Klebanov and A.~M.~Polyakov,
  Phys.\ Lett.\  B {\bf 428}, 105 (1998)
  [arXiv:hep-th/9802109].

  \bibitem{lm4}
O.~Lunin and S.~D.~Mathur,
  ``AdS/CFT duality and the black hole information paradox,''
  Nucl.\ Phys.\  B {\bf 623}, 342 (2002)
  [arXiv:hep-th/0109154];



\bibitem{fuzzballs}
  O.~Lunin, J.~M.~Maldacena and L.~Maoz,
  hep-th/0212210;
 S.~D.~Mathur,
  Fortsch.\ Phys.\  {\bf 53}, 793 (2005)
  [arXiv:hep-th/0502050];\\
  V.~Jejjala, O.~Madden, S.~F.~Ross and G.~Titchener,
  ``Non-supersymmetric smooth geometries and D1-D5-P bound states,''
  Phys.\ Rev.\  D {\bf 71}, 124030 (2005)
  [arXiv:hep-th/0504181];\\
V.~Balasubramanian, E.~G.~Gimon and T.~S.~Levi,
  JHEP {\bf 0801}, 056 (2008)
  [arXiv:hep-th/0606118];\\
I.~Bena and N.~P.~Warner,
  Lect.\ Notes Phys.\  {\bf 755}, 1 (2008)
  [arXiv:hep-th/0701216];\\
 K.~Skenderis and M.~Taylor,
  Phys.\ Rept.\  {\bf 467}, 117 (2008)
  [arXiv:0804.0552 [hep-th]];\\
  I.~Bena, S.~Giusto, E.~J.~Martinec, R.~Russo, M.~Shigemori, D.~Turton and N.~P.~Warner,
  Phys.\ Rev.\ Lett.\  {\bf 117}, no. 20, 201601 (2016)
  doi:10.1103/PhysRevLett.117.201601
  [arXiv:1607.03908 [hep-th]].


\bibitem{radiation}
  V.~Cardoso, O.~J.~C.~Dias, J.~L.~Hovdebo and R.~C.~Myers,
  ``Instability of non-supersymmetric smooth geometries,''
  Phys.\ Rev.\  D {\bf 73}, 064031 (2006)
  [arXiv:hep-th/0512277];
  B.~D.~Chowdhury and S.~D.~Mathur,
  ``Radiation from the non-extremal fuzzball,''
  Class.\ Quant.\ Grav.\  {\bf 25}, 135005 (2008)
  [arXiv:0711.4817 [hep-th]].

       
   \bibitem{tunnel}
  S.~D.~Mathur,
  arXiv:0805.3716 [hep-th];
  S.~D.~Mathur,
  Int.\ J.\ Mod.\ Phys.\  D {\bf 18}, 2215 (2009)
  [arXiv:0905.4483 [hep-th]].
  
\bibitem{kraus}
  P.~Kraus and S.~D.~Mathur,
  Int.\ J.\ Mod.\ Phys.\ D {\bf 24}, no. 12, 1543003 (2015)
  doi:10.1142/S0218271815430038
  [arXiv:1505.05078 [hep-th]].
  
  
  \bibitem{puhm}
  I.~Bena, D.~R.~Mayerson, A.~Puhm and B.~Vercnocke,
  JHEP {\bf 1607}, 031 (2016)
  doi:10.1007/JHEP07(2016)031
  [arXiv:1512.05376 [hep-th]].

\bibitem{model} 
  S.~D.~Mathur,
  Int.\ J.\ Mod.\ Phys.\ D {\bf 25}, no. 12, 1644018 (2016)
  doi:10.1142/S0218271816440181
  [arXiv:1609.05222 [hep-th]].

   
\bibitem{buchdahl} 
  H.~A.~Buchdahl,
  Phys.\ Rev.\  {\bf 116}, 1027 (1959).
  doi:10.1103/PhysRev.116.1027


\bibitem{witten} 
  E.~Witten,
  Nucl.\ Phys.\ B {\bf 195}, 481 (1982).
  
\bibitem{sen} 
  A.~Sen,
  JHEP {\bf 9710}, 002 (1997)
  doi:10.1088/1126-6708/1997/10/002
  [hep-th/9708002].
  
\bibitem{beyond} 
  S.~D.~Mathur,
  Annals Phys.\  {\bf 327}, 2760 (2012)
  doi:10.1016/j.aop.2012.05.001
  [arXiv:1205.0776 [hep-th]].
  
\bibitem{gibbonswarner} 
  G.~W.~Gibbons and N.~P.~Warner,
  Class.\ Quant.\ Grav.\  {\bf 31}, 025016 (2014)
  doi:10.1088/0264-9381/31/2/025016
  [arXiv:1305.0957 [hep-th]].
  
\bibitem{universe} 
  S.~D.~Mathur,
  Int.\ J.\ Mod.\ Phys.\ D {\bf 12}, 1681 (2003)
  doi:10.1142/S0218271803004031
  [hep-th/0305204].
  
  
  
\bibitem{dasjevicki} 
  S.~R.~Das and A.~Jevicki,
  Mod.\ Phys.\ Lett.\ A {\bf 5}, 1639 (1990).
  doi:10.1142/S0217732390001888
  
\bibitem{jevickinonpert} 
  A.~Jevicki,
  Nucl.\ Phys.\ B {\bf 376}, 75 (1992).
  doi:10.1016/0550-3213(92)90068-M




  
\bibitem{pol} 
  J.~Polchinski,
  hep-th/9411028.
  
    
  \bibitem{jevicki}
A.~Jevicki, \, {\it unpublished}.


  
\bibitem{kms} 
  J.~L.~Karczmarek, J.~M.~Maldacena and A.~Strominger,
  JHEP {\bf 0601}, 039 (2006)
  doi:10.1088/1126-6708/2006/01/039
  [hep-th/0411174].
  
\bibitem{folds} 
  S.~R.~Das and S.~D.~Mathur,
  Phys.\ Lett.\ B {\bf 365}, 79 (1996)
  doi:10.1016/0370-2693(95)01307-5
  [hep-th/9507141].
  
\bibitem{vilenkin} 
  J.~Garriga, S.~Kanno, M.~Sasaki, J.~Soda and A.~Vilenkin,
  JCAP {\bf 1212}, 006 (2012)
  doi:10.1088/1475-7516/2012/12/006
  [arXiv:1208.1335 [hep-th]];
  M.~B.~Frob, J.~Garriga, S.~Kanno, M.~Sasaki, J.~Soda, T.~Tanaka and A.~Vilenkin,
  JCAP {\bf 1404}, 009 (2014)
  doi:10.1088/1475-7516/2014/04/009
  [arXiv:1401.4137 [hep-th]].
  
\bibitem{deathpayne} 
  P.~D.~D'Eath and P.~N.~Payne,
  Phys.\ Rev.\ D {\bf 46}, 658 (1992).
  doi:10.1103/PhysRevD.46.658

\bibitem{mtflaw} 
  S.~D.~Mathur and D.~Turton,
  Nucl.\ Phys.\ B {\bf 884}, 566 (2014)
  doi:10.1016/j.nuclphysb.2014.05.012
  [arXiv:1306.5488 [hep-th]].
  
\bibitem{virmani} 
  A.~J.~Amsel, D.~Marolf and A.~Virmani,
  JHEP {\bf 0804}, 025 (2008)
  doi:10.1088/1126-6708/2008/04/025
  [arXiv:0712.2221 [hep-th]].
  
\bibitem{amps} 
  A.~Almheiri, D.~Marolf, J.~Polchinski, J.~Sully and ,
  JHEP {\bf 1302}, 062 (2013)
  [arXiv:1207.3123 [hep-th]].

  
  \bibitem{thooft}
 G.~'t Hooft,
  ``The Holographic principle: Opening lecture,''
  hep-th/0003004;


  
  \bibitem{susskind}
  L.~Susskind, L.~Thorlacius, J.~Uglum,
  ``The Stretched horizon and black hole complementarity,''
  Phys.\ Rev.\  {\bf D48}, 3743-3761 (1993).
  [hep-th/9306069];
    L.~Susskind,
  ``String theory and the principles of black hole complementarity,''
  Phys.\ Rev.\ Lett.\  {\bf 71}, 2367-2368 (1993).
  [hep-th/9307168];
  L.~Susskind,
  ``The World As A Hologram,''
  J.\ Math.\ Phys.\  {\bf 36}, 6377 (1995)
  [arXiv:hep-th/9409089];
    D.~A.~Lowe, J.~Polchinski, L.~Susskind {\it et al.},
  ``Black hole complementarity versus locality,''
  Phys.\ Rev.\  {\bf D52}, 6997-7010 (1995).
  [hep-th/9506138].


  
\bibitem{plumberg} 
  S.~D.~Mathur and C.~J.~Plumberg,
  ``Correlations in Hawking radiation and the infall problem,''
  JHEP {\bf 1109}, 093 (2011)
  [arXiv:1101.4899 [hep-th]].




\bibitem{bitmodel} 
  S.~D.~Mathur,
  arXiv:1506.04342 [hep-th].
  
  




  
\bibitem{cool} 
  J.~Maldacena and L.~Susskind,
  Fortsch.\ Phys.\  {\bf 61}, 781 (2013)
  doi:10.1002/prop.201300020
  [arXiv:1306.0533 [hep-th]].

\bibitem{pr} 
  K.~Papadodimas and S.~Raju,
  JHEP {\bf 1310}, 212 (2013)
  doi:10.1007/JHEP10(2013)212
  [arXiv:1211.6767 [hep-th]].


\bibitem{giddings} 
  S.~B.~Giddings,
  ``Nonviolent information transfer from black holes: a field theory parameterization,''
  arXiv:1302.2613 [hep-th].

\bibitem{hps} 
  S.~W.~Hawking, M.~J.~Perry and A.~Strominger,
  Phys.\ Rev.\ Lett.\  {\bf 116}, no. 23, 231301 (2016)
  doi:10.1103/PhysRevLett.116.231301
  [arXiv:1601.00921 [hep-th]].


\bibitem{cdt} 
  J.~Ambjorn, J.~Jurkiewicz and R.~Loll,
  Phys.\ Rev.\ Lett.\  {\bf 93}, 131301 (2004)
  doi:10.1103/PhysRevLett.93.131301
  [hep-th/0404156].
  
\bibitem{sorkin} 
  L.~Bombelli, J.~Lee, D.~Meyer and R.~Sorkin,
  Phys.\ Rev.\ Lett.\  {\bf 59}, 521 (1987).
  doi:10.1103/PhysRevLett.59.521
  

\end{thebibliography}
\end{document}